\documentclass[12pt]{article}
\usepackage{amsmath,amssymb}
\usepackage{latexsym,graphicx}

\def\nn{\nonumber\\}

\def\6#1{{\underline{#1}}}
\def\m6#1{{\underline{#1}\,}}

\newdimen\Tdim
\def\ispan{{\setbox0=\hbox{i}%
\Tdim\ht0\advance\Tdim\dp0\rule[-\dp0]{0pt}{\Tdim}}}
\def\jspan{{\setbox0=\hbox{j}%
\Tdim\ht0\advance\Tdim\dp0\rule[-\dp0]{0pt}{\Tdim}}}
\def\Tspan#1{{\setbox0=\hbox{#1}%
\Tdim\ht0\advance\Tdim\dp0\advance\Tdim.55ex\rule[-\dp0]{0pt}{\Tdim}\box0}}

\def\be{\begin{eqnarray}}
\def\ben{\begin{eqnarray*}}
\def\ee{\end{eqnarray}}
\def\een{\end{eqnarray*}}
\def\Tr{{\rm Tr}}

\def\p{\partial}

\def\D{\mathcal{D}}

\def\=:{=\hspace{-.7em}\raisebox{1.1ex}{.}\hspace{.1em}\raisebox{-0.2ex}{.} }
\newcommand{\ibf}[1]{\mbox {\boldmath{$ #1 $}}}
\newcommand{\NF}{N_{\rm F}}
\newcommand{\NC}{N_{\rm C}}
\newcommand{\hs}[1]{\hspace{#1 mm}}

\newcommand {\beq}{\begin{eqnarray}}
\newcommand {\eeq}{\end{eqnarray}}
\newcommand {\non}{\nonumber\\}

\makeatletter
\@addtoreset{equation}{section}

\renewcommand{\thefootnote}{\fnsymbol{footnote}}
\makeatother

\makeatletter
\newcommand{\thetablename}{Table}
\def\fnum@table{\thetablename\ \thetable}
\makeatother

\begin{document}
\thispagestyle{empty}
\begin{flushright}
UT-Komaba/06-14, TIT/HEP-563, \\
{\tt hep-th/0612003} \\
Nov, 2006 \\
\end{flushright}
\vspace{3mm}
\begin{center}
{\LARGE \bf 
Effective Action of Domain Wall Networks
} \\ 
\vspace{5mm}

{\normalsize\bfseries
Minoru~Eto$^\dagger$, 
Toshiaki~Fujimori, 
Takayuki Nagashima, \\
Muneto~Nitta$^{\dagger\dagger}$,
Keisuke~Ohashi, 
and
 Norisuke~Sakai}
\footnotetext{
e-mail~addresses: \tt
meto@hep1.c.u-tokyo.ac.jp;
fujimori,
nagashi@th.phys.titech.ac.jp;\\
nitta@phys-h.keio.ac.jp;
keisuke,
nsakai@th.phys.titech.ac.jp
}

\vskip 1.5em
{\it Department of Physics, Tokyo Institute of
Technology, Tokyo 152-8551, Japan}\\
$^\dagger$ {\it Institute of Physics, University of Tokyo, Komaba 3-8-1,
Tokyo 153-8902, Japan}\\
$^{\dagger\dagger}$ 
{\it Department of Physics, Keio University, Hiyoshi, Yokohama,
Kanagawa 223-8521, Japan}
\vspace{12mm}

\abstract{

$U(\NC)$ gauge theory with $\NF$ fundamental scalars admits 
BPS junctions of domain walls. 
When the 
networks/webs 
of these 
walls 
contain loops, their 
size moduli give localized massless modes. 
We construct K\"ahler potential of their effective action. 
In the large size limit K\"ahler metric is well 
approximated by kinetic energy of walls and junctions, 
which is understood in terms of tropical geometry. 
K\"ahler potential can be expressed in terms of 
hypergeometric functions which are useful to understand 
small size behavior.
Even when the loop shrinks, 
the metric is regular with positive curvature. 
Moduli space 
of a single triangle loop
has a geometry between a cone and a cigar. 
}

\end{center}

\vfill
\newpage
\setcounter{page}{1}
\setcounter{footnote}{0}
\renewcommand{\thefootnote}{\arabic{footnote}}

\section{Introduction}\label{sec:intro}

Network of topological defects (solitons) is 
ubiquitous in various area of physics.
For instance in the early Universe 
a network of domain walls 
or cosmic strings is inevitable 
at a phase transition via the Kibble mechanism \cite{Kibble:1976sj,VS}. 
A network of domain walls \cite{cosmology-wall} was proposed as
a candidate of dark matter/energy \cite{dark-energy}.
Networks of domain walls appear also in 
several situations in condensed matter physics 
(see, e.g., \cite{cond-mat}).
Dynamics of these networks have been examined 
by numerical simulations so far 
because analytic solutions were lacking.\footnote{
However some analytic solutions are known 
in integrable systems (coupled KP's) \cite{junction-cKP}.
}
In supersymmetric field theory, 
junctions of domain walls \cite{Abraham:1990nz} 
were found to be 1/4 Bogomol'nyi-Prasad-Sommerfield (BPS) 
states preserving only a quarter of supersymmetry  
\cite{Gibbons:1999np}.
Since then many works about BPS or nearly BPS networks 
of domain walls 
have been done during these years 
\cite{Oda:1999az}--\cite{network}.  
However most works relied on qualitative treatments 
or numerical simulations 
except for a few works on 
exact solutions of a single junction of walls 
\cite{Oda:1999az,Kakimoto:2003zu}. 
Recently the most general exact solutions of networks (webs) of 
domain walls have been found in a supersymmetric 
Abelian/non-Abelian gauge field theory 
\cite{Eto:2005cp,Eto:2005fm,Eto:2005mx}. 
These solutions contain full moduli of a network with arbitrary numbers 
of loops and external legs of walls.
The purpose of this paper is to construct effective field 
theory of the networks of domain walls
by determining the moduli space metric. 
It will be possible to describe 
dynamics of these networks 
by the moduli space (geodesic) 
approximation \`a la Manton \cite{Manton:1981mp} 
which remains as a future problem.

Our model 
is a $U(N_{\rm C})$ gauge theory 
coupled to $N_{\rm F}$ Higgs fields 
in the fundamental representation 
which has been recently studied extensively 
and has been found to 
allow many kinds of solitons.
If the number $N_{\rm F}$ of flavors is equal to or larger 
than the number of colors of $U(N_{\rm C})$ gauge group, 
$N_{\rm F} \geq N_{\rm C}$, 
vacua are in the Higgs phase.  
The model at the critical coupling 
(with a gauge coupling constant 
and a Higgs self-coupling constant being equal)
admits ${\cal N}=2$ supersymmetric extension,  
and solitons become 
BPS states and are the most stable among configurations 
with the same topological charges. 
Domain walls and vortex-strings 
are fundamental BPS solitons in the Higgs phase, 
preserving/breaking a half of supersymmetry,  
and are called 1/2 BPS states.  
Parallel multiple domain wall solutions can exist
when Higgs mass parameters are real and non-degenerate
and vacua are in the Higgs phase with 
disconnected branches ($N_{\rm F} > N_{\rm C}$). 
By introducing the method of the moduli matrix 
\cite{Eto:2006pg,rev-modulimatrix} 
those solutions were constructed 
\cite{Isozumi:2004jc,Isozumi:2004va,Eto:2004vy,Eto:2005wf} 
with extending earlier works of $U(1)$ gauge theory 
\cite{U(1)walls}--\cite{Hanany:2005bq}. 
Domain walls with non-Abelian flavor symmetry 
were also constructed in theory with 
partially degenerate Higgs masses \cite{Shifman:2003uh,Eto:2005cc}.
The model with massless Higgs fields 
admits parallel multiple non-Abelian 
vortex-strings \cite{Hanany:2003hp}, \cite{vortices}--\cite{Eto:2006db}. 
They are local strings \cite{ANO} for $N_{\rm F} = N_{\rm C}$ 
or semi-local strings \cite{Vachaspati:1991dz} for $N_{\rm F} > N_{\rm C}$. 
The moduli matrix enables us to obtain the formal solutions of them 
and their moduli space \cite{Eto:2005yh,Eto:2006mz,Eto:2006pg,Eto:2006cx}.  
In both cases of domain walls and vortices, 
the half of BPS equations is 
solved in terms of the moduli matrix, 
and the rest of them 
is rewritten as a second order differential equation 
in terms of a gauge invariant quantity $\Omega$. 
This equation, called the master equation, 
is expected to admit the unique solution.\footnote{
The uniqueness was rigorously proven in various cases: 
vortices \cite{Taubes:1979tm}, domain walls \cite{Sakai:2005kz} and 
domain wall webs \cite{Eto:2005cp} in Abelian gauge theory. 
In the first two cases, the existence was also shown.
}
Effective K\"ahler potential of these 1/2 BPS solitons 
was constructed as an integration of a function of $\Omega$ 
in the superfield formalism \cite{Eto:2006uw}. 
The integration can be explicitly
performed to obtain the explicit 
K\"ahler potential/metric for several cases, 
so we can discuss the dynamics of these solitons. 
The dynamics of domain walls in $U(1)$ gauge theory 
was discussed \cite{Tong:2002hi,Tong:2003ik,Isozumi:2003rp}: 
for example the moduli space of 
a double wall in $U(1)$ gauge theory 
with $N_{\rm F}=3$ 
is ${\bf C}$ 
with the cigar metric\footnote{
This is the 
modulus corresponding to the relative distance 
and the associated phase. 
There is also another complex modulus 
corresponding to the 
center of mass of two domain walls and the associated phase, 
which has a trivial flat metric. 
}
(2d Euclidean black hole). 
The moduli space of single non-Abelian vortex 
was found to be ${\bf C}P^{N-1}$ for 
$N_{\rm F} = N_{\rm C} =N$ \cite{Hanany:2003hp}.
The dynamics of two non-Abelian vortex-strings 
has been studied recently in order to describe 
colliding two vortex-strings with angle, 
with resulting in their reconnection \cite{Eto:2006db}. 
However the dynamics of composite solitons 
was not discussed so far.

In this paper, by generalizing the result of \cite{Eto:2006uw}, 
we obtain a 
general formula
of the effective K\"ahler potential 
of the network/web of domain walls 
in the $U(N_{\rm C})$ gauge theory coupled to 
$N_{\rm F}$ Higgs scalar fields with complex and non-degenerate masses.
This system of web of domain walls contains 
non-normalizable modes associated with the modes on walls 
at asymptotic infinity. 
Integration over non-normalizable modes brings 
a divergence of the K\"ahler potential  
and therefore these non-normalizable modes 
are to be fixed by boundary conditions.
On the other hand, 
apparent infra-red divergences 
appearing in the integration of 
normalizable modes 
can be eliminated by K\"ahler transformations. 
We thus can consider only the normalizable 
modes as effective fields on the web. 
We have already pointed out that 
only possible normalizable modes come from 
the size and associated 
phase of a loop \cite{Eto:2005cp}. 
As the simplest example 
we discuss a single triangle loop of domain walls 
in this paper.
Moduli of the loop size and its associated $U(1)$ phase are 
combined to give one complex modulus $\phi$.
We explicitly perform integrations over two codimensions 
in the strong coupling limit $g^2\to \infty$ 
to obtain the K\"ahler potential 
expressed in terms of hypergeometric functions. 
This is a Taylor expansion with respect to $|\phi|$ 
and is useful to understand 
small size behavior of the loop. 
In the large size limit the K\"ahler metric is well 
approximated by kinetic energy of walls and junctions, 
which is understood in terms of tropical geometry. 
The metric is regular with positive curvature everywhere, and 
no singularity appears even when the loop shrinks.
The geometry is between a cone and a cigar.

Apart from the dynamics of BPS solitons, 
there is another merit to construct an explicit metric on 
a moduli space of BPS solitons. 
When one obtains an explicit metric of the moduli space 
of 1/2 BPS solitons and the effective theory on them as 
a sigma model with the moduli space metric as its target 
space, 1/4 BPS composite solitons often can be 
described as another 1/2 BPS soliton solutions 
in terms of the effective theory on the host solitons. 
For instance, by using the ${\bf C}P^{N-1}$ model 
as the effective theory 
on a single vortex, 
confined monopoles can be 
realized as kinks inside the vortex \cite{Tong:2003pz} 
whereas instantons can be realized as ${\bf C}P^{N-1}$-lumps inside 
the vortex \cite{Eto:2004rz}.\footnote{Even non-BPS solitons can be 
realized in the same way. 
Instantons can be realized as 
Skyrmions in domain walls in five dimensions \cite{Eto:2005cc}.
}
With proceeding in this way, 
one may be able to 
construct 1/8 BPS solitons \cite{Lee:2005sv,Eto:2005sw} 
once one obtains an explicit metric on
the moduli space of 1/4 BPS solitons. 
The metric on the wall web obtained in this paper 
may 
be applied to this direction.

There exist 1/4 BPS composite states made of parallel 
domain walls attached or stretched by vortex-strings 
\cite{Gauntlett:2000de,Isozumi:2004vg}. 
In strong gauge coupling limit, 
all solutions were obtained exactly
\cite{Isozumi:2004vg}. 
These resemble with the Hanany-Witten brane configurations 
in the type IIA/B string theory \cite{brane-config}. 
The type IIB string theory admits
$(p,q)$ string/5-brane webs \cite{string-webs,1/4dyon,5-brane} 
which are also 1/4 BPS states.
Low energy dynamics of these webs 
is described by $d=5$ gauge theory. 
Corresponding example of field theory 
is a domain wall web \cite{Eto:2005cp,Eto:2005fm,Eto:2005mx} 
which will be discussed in this paper. 
It has been shown in \cite{Eto:2005cp,Eto:2005sw}
that the dynamics is described by a $d=2$, ${\cal N}=(2,0)$ 
nonlinear sigma model \cite{Hull:1985jv,Witten:2005px}.
Pursuing the similarity between string theory branes 
and field theory branes is another motivation 
of the study in this paper.

This paper is organized as follows. 
In Sec.~\ref{sec:model}, we briefly review the basic properties of webs of walls. In Sec.~\ref{sec:effective}, we construct an effective theory on the webs of walls. In \ref{subsec:general}, we give a general form of the effective action and show that the effective theory is described by a nonlinear sigma model whose target space is the moduli space with a K\"ahler metric. In \ref{subsec:1-loop}, we analyze the case of a single triangle loop and in \ref{subsec:hypergeometric} we show that the K\"ahler potential can be obtained as a power series of the moduli parameter. In Sec.~\ref{sec:tropical}, we examine the large size behavior of the single triangle loop by taking a special limit, which we call ``tropical limit". In \ref{subsec:parallel}, we consider the asymptotic behavior of the K\"ahler potential in the case of 1/2 BPS parallel walls to illustrate what the tropical limit is. In \ref{subsec:troloop}, applying the tropical limit in the case of the single triangle loop, w
 e obtain an asymptotic metric of the moduli space and in \ref{subsec:finite} we also evaluate the contributions from junction charges. In Sec.~\ref{sec:understand}, we show that the asymptotic metric can be read from kinetic energy of the constituent walls of the loop and the junction charges. Sec.~\ref{sec:discussion} is devoted to conclusion and discussion.




 \section{Networks/Webs of Walls}\label{sec:model}

Let us here briefly present our model admitting the 1/4 BPS webs of domain walls 
(see \cite{Eto:2006pg} for a review).
Our model is 3+1 dimensional $\mathcal{N}=2$ supersymmetric $U(\NC)$ gauge theory with $\NF(>\NC)$
massive hypermultiplets in the fundamental representation. 
Here the bosonic components in the vector multiplet are gauge fields 
$W_M~(M=0,1,2,3)$, the real scalar fields $\Sigma_\alpha~(\alpha=1,2)$
in the adjoint representation, and those in the hypermultiplet are the $SU(2)_R$
doublets of the complex scalar fields $H^i~(i=1,2)$, which we express as $\NC \times \NF$ matrices. 
The bosonic part of the Lagrangian is given by
\beq
\mathcal{L} &=& \Tr \left[-\frac{1}{2g^2}F_{MN}F^{MN}+\frac{1}{g^2} \sum_{\alpha=1}^2
\mathcal D_M \Sigma_\alpha \mathcal D^M \Sigma_\alpha
+\mathcal D_M H^i (\mathcal D^M H^i)^\dag \right] - V, \label{eq:lag}\\
V &=& \Tr \left[ \frac{1}{g^2} \sum_{a=1}^3 (Y^a)^2 
+ \sum_{\alpha=1}^2(H^iM_\alpha-\Sigma_\alpha H^i)(H^iM_\alpha-\Sigma_\alpha H^i)^\dag
-\frac{1}{g^2}[\Sigma_1,\Sigma_2]^2 \right],
\label{eq:pot}
\eeq
where we have defined 
$Y^a \equiv \frac{g^2}{2}\left(c^a \mathbf 1_{\NC} - {(\sigma^a)^j}_i H^i(H^j)^\dagger \right)$ 
with $c^a$ an $SU(2)_R$  triplet of the Fayet-Iliopoulos (FI) parameters. 
In the following, we choose the FI parameters as $c^a=(0,0,c>0)$ 
by using $SU(2)_R$ rotation without loss of generality. 
Here we use the space-time metric $\eta_{MN}=\text{diag}(+1,-1,-1,-1)$ 
and $M_\alpha$ are real diagonal mass matrices, $M_1=\text{diag}
(m_1,m_2,\cdots,m_{\NF}),~M_2=\text{diag}(n_1,n_2,\cdots,n_{\NF})$. 
The covariant derivatives are defined as 
$\mathcal D_M \Sigma =\partial_M \Sigma+ i[W_M,\Sigma],~\mathcal D_M H^i=(\partial_M+iW_M)H^i$, and the
field strength is defined as 
$F_{MN}=-i[\mathcal D_M,\mathcal D_N]=\partial_M W_N-\partial_N W_M +i[W_M,W_N]$. 

If we turn off all the mass parameters, the vacuum manifold is
the cotangent bundle over the complex Grassmannian $T^\ast Gr_{\NF,\NC}$ 
\cite{Lindstrom:1983rt}. 
Once the mass parameters $m_A + i n_A,~(A=1,\cdots \NF)$ 
are turned on and chosen to be fully nondegenerate ($m_A+in_A \neq m_B + in_B$ for $A\neq B$),
the almost all points of the vacuum manifold 
are lifted and only ${}_{\NF} C_{\NF}=\NF!/\left[\NC!(\NF-\NC)!\right]$ discrete points 
on the base manifold $Gr_{\NF,\NC}$ are left to be the supersymmetric vacua \cite{Arai:2003tc}. 
Each vacuum is characterized by a set of $\NC$ different indices $
\ \left< A_1,\cdots,A_{\NC}\right> $ such that
$1\leq A_1 < \cdots < A_{\NC} \leq \NF$.
In these vacua, the vacuum expectation values are determined as
\beq
&\left<H^{1rA}\right> = \sqrt{c} \delta^{A_r}{}_A,\hs{5} \left<H^{2rA}\right> = 0,&\non
& \left<\Sigma\right> = {\rm diag} \left(m_{A_1}+in_{A_1},\cdots,m_{A_{\NC}}+in_{A_{\NC}}\right), &
\label{vacuum}
\eeq
where 
$r$ is color index running from 1 to $\NC$, the flavor index $A$ runs from 1 to $\NF$ and
$\Sigma$ is the complex adjoint scalar defined by $\Sigma \equiv \Sigma_1 + i \Sigma_2$.


The 1/4 BPS equations for webs of walls interpolating the 
discrete vacua (\ref{vacuum}) can be obtained 
by usual Bogomolny completion \cite{Eto:2005cp,Eto:2005fm} 
of the energy density as 
\begin{align}
&F_{12}=i[\Sigma_1,\Sigma_2],~~~~\mathcal D_1\Sigma_2=\mathcal D_2\Sigma_1,~~~~
\mathcal D_1\Sigma_1+\mathcal D_2\Sigma_2=Y^3,\label{eq:BPS2} \\
&\mathcal D_1H^1=H^1M_1-\Sigma_1H^1,~~~~\mathcal D_2H^1=H^1M_2-\Sigma_2H^1. \label{eq:BPS1}
\end{align}
Here we consider static configurations which are 
independent of $x_3$, so we set $\partial_0=\partial_3=0$ and $W_0=W_3=0$. 
Furthermore, we take $H^2=0$ because it always vanishes 
for the 1/4 BPS solutions. 
Let us solve the 1/4 BPS equations (\ref{eq:BPS2}) and 
(\ref{eq:BPS1})~\cite{Eto:2005cp,Eto:2005fm}.
Firstly (\ref{eq:BPS1}) is solved by
\begin{align}
H^1=S^{-1}(x^1,x^2)H_0e^{M_1x^1+M_2x^2},\quad 
\Sigma_\alpha+iW_\alpha=S^{-1}(x^1,x^2)\partial_\alpha S(x^1,x^2).
\label{eq:solBPS}
\end{align}
Here $H_0$ is an $\NC \times \NF$ constant complex matrix of rank $\NC$, 
and contains all the moduli parameters of solutions. 
The matrix valued quantity $S(x^1,x^2) \in GL(\NC,{\bf C})$ is determined by 
the remaining equation (\ref{eq:BPS2}) as
we will see shortly.
The moduli matrices related by the following 
$V$-transformation are physically equivalent since they do not change the physical configuration:
\begin{align}
H_0 \rightarrow VH_0,~~~~S(x^1,x^2) \rightarrow VS(x^1,x^2),~~~~V \in GL(\NC, {\bf C}).
\end{align}
Secondly the first two equations in (\ref{eq:BPS2}) give 
an integrability condition for the two operators
${\cal D}_\alpha + \Sigma_\alpha\ (\alpha=1,2)$, so they are automatically satisfied.
Finally the last equation in (\ref{eq:BPS2}) can be converted, 
by using a gauge invariant quantity 
\begin{align}
 \Omega \equiv SS^\dag, 
\end{align}
to the following equation:
\begin{equation}
\frac{1}{cg^2}\bigl(\partial_\alpha(\Omega^{-1}\partial_\alpha \Omega)\bigl)=
\ibf{1}_{\NC}-\Omega^{-1}\Omega_0,
\label{eq:master}
\end{equation}
\begin{equation}
\Omega_0 \equiv \frac{1}{c}H_0 e^{2(M_1x^1+M_2x^2)} H_0^\dag .
\label{eq:omega0}
\end{equation}
The solution $\Omega$ of the master equation should 
approach $\Omega_0$ near vacuum regions. 
This equation is called the master equation for webs of walls. 
It determines $S$ for a given moduli matrix $H_0$ up to the gauge symmetry 
and then the physical fields can be obtained through (\ref{eq:solBPS}).

Solutions of the 1/4 BPS equations saturate the Bogomol'nyi energy bound
\begin{align}
\mathcal{E} 
\geq \mathcal{Y}+\mathcal{Z}_1+\mathcal{Z}_2+\partial_\alpha J_\alpha,
  \label{eq:Bogomolny}
\end{align}
where the central (topological) charge densities which
characterize the solutions are of the form
\begin{align}
\mathcal{Y}=\frac{2}{g^2}\partial_\alpha 
\text{Tr}(\epsilon^{\alpha \beta}\Sigma_2 \mathcal D_\beta \Sigma_1),~~~~
\mathcal{Z}_1=c \partial_1 \text{Tr} \Sigma_1,~~~~\mathcal{Z}_2=c \partial_2 \text{Tr} \Sigma_2.
\end{align}
The topological charges are defined by
$
T_w \equiv \int d^2 x \left( \mathcal Z_1 + \mathcal Z_2 \right)
$ and
$
Y \equiv \int d^2 x \, \mathcal Y.
$
Here $T_w$ corresponds to the 
energy of domain walls 
and $Y$ corresponds to the energy 
of the domain wall junction. 
Since energy of domain walls means tension times the length of the walls, this quantity is divergent. 
On the other hand $Y$ has a finite value, 
and we call this charge as the junction charge or the Hitchin charge.
Note that the integration of the fourth term
$
\partial_\alpha J_\alpha=\partial_\alpha \text{Tr}[H^1(M_\alpha H^{1\dag}-H^{1\dag} \Sigma_\alpha)]
$
in (\ref{eq:Bogomolny}) 
does not contribute to the topological charges. The above energy bound
can be rewritten in terms of $\Omega$ as
\begin{eqnarray}
{\cal E}_{\rm BPS} = 
\frac{2}{g^2}\Tr\left[\epsilon^{\alpha\beta}
\partial_\alpha\left(\partial_2\Omega\Omega^{-1}\right)
\partial_\beta\left(\partial_1\Omega\Omega^{-1}\right)
\right] + 
\left(\frac{c}{2} - \frac{1}{2g^2}\partial_\alpha^2\right)
\partial_\alpha^2\log\det\Omega.
\label{energy}
\end{eqnarray}

\begin{figure}[ht]
\begin{center}
\includegraphics[height=5cm]{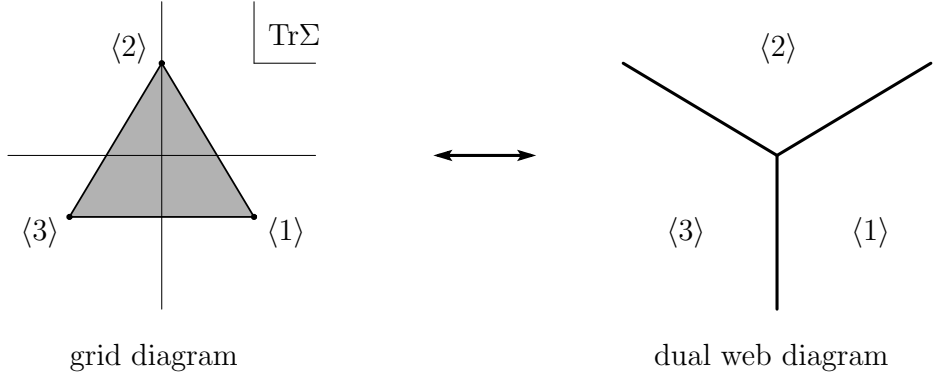}
\caption{The 3-pronged junction in the Abelian gauge theory. 
The left figure shows the grid diagram and the right shows 
its dual web diagram in the real spacial ($x^1$-$x^2$) plane.}
\label{fig:grid}
\end{center}
\end{figure}
\begin{figure}[h]
\begin{center}
\includegraphics[width=60mm]{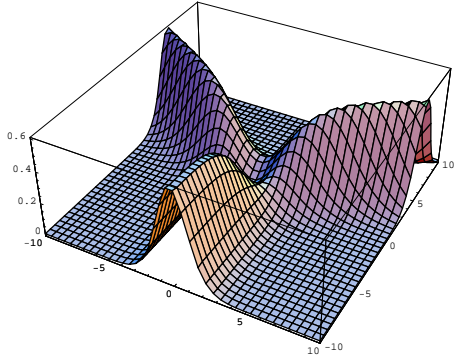}
\caption{The binding energy at the junction 
point:~The energy density is numerically evaluated for 
the moduli matrix 
$ H_0 e^{\vec{m} \cdot \vec{x}} 
= \left( e^{x^2},\, e^{\sqrt{3}x^1/2-x^2/2},
\,e^{ - \sqrt{3}x^1/2-x^2/2}\right)$, 
gauge coupling $g=1$ and FI parameter $c=1$. 
The hole of the energy density at the junction point can 
be regarded as binding energy of the three walls.}
\label{fig:junction}
\end{center}
\end{figure}

There is a useful diagram to understand the structure of webs of walls, which
is called the grid diagram \cite{Eto:2005cp,Eto:2005fm}.
The grid diagram is a convex in the complex plane $\Tr \Sigma\ (\Sigma \equiv \Sigma_1+i\Sigma_2)$.
The $_{\NF} C_{\NC}$ vacuum points (\ref{vacuum}) correspond to
the vertices of the convex which are plotted at 
$\Tr\left<\Sigma\right> = \sum_{r=1}^{\NC}m_{\left<A_r\right>} + i\sum_{r=1}^{\NC}n_{\left<A_r\right>}$.
Then, each edge connecting two vertices corresponds to a domain wall
interpolating the two vacua\footnote{
Only two vacua such as have only one different element in 
their labels like $\left<\ \underline{\dots}\ A\right>$ 
and $\left<\ \underline{\dots}\ B\right>$ can be connected 
while two with $\left<\ \underline{\dots}\ AB\right>$ 
and $\left<\ \underline{\dots}\ CD\right>$ are forbidden 
to be connected.
},
and each triangle corresponds to a 3-pronged domain wall 
junction. 
An example of the grid diagram for a 3-pronged junction 
with $N_{\rm F}=3$ and $N_{\rm C}=1$ 
is depicted in Fig.\ref{fig:grid}.
One can easily read physical informations about domain 
walls and junctions from the grid diagram:
The tension 
of the domain wall is proportional to the length 
of the corresponding edge of the grid diagram while 
the junction charge is to the area of the triangle. 
We have found previously that the topological charge of 
the Abelian junction gives a negative contribution to the 
energy of the configuration which can be interpreted as a 
binding energy of walls. We show the energy density of
the Abelian junction in Fig.(\ref{fig:junction}).
One should note that the three vacua 
 $\langle 1\rangle$, $\langle 2\rangle$ 
and $\langle 3\rangle$ are ordered 
{\it counterclockwisely}. 
On the other hand,  non-Abelian junction appears in the 
case of $U(2)$ gauge theory ($N_{\rm C}=2$) and 
$N_{\rm F}=3$ with the same masses for hypermultiplets as 
in the above abelian theory. 
Then there appear the three vacua $\langle 12\rangle$, 
$\langle 23\rangle$and $\langle 31\rangle$ in the grid 
diagram which is congruent to the grid diagram of the Abelian case in Fig.\ref{fig:grid} : 
These two grid diagrams coincide when one of them is 
rotated by the angle $\pi$. 
Therefore tension of domain walls
also coincides.
However,
the orientation for the non-Abelian junction 
is opposite to the Abelian junction, namely it is 
{\it clockwise}.
This feature makes the $Y$-charge of the non-Abelian 
junction to contribute positively to the energy
with the same magnitude.
At the non-Abelian junction points an interesting system 
which is called the Hitchin system is realized.
This positive $Y$-charge has been understood as 
the Hitchin charge of the Hitchin system \cite{Eto:2005fm}. 


In order to extract concrete informations from the moduli 
matrix $H_0$, it is useful to define the weight 
${\cal W}^{\left<A_r\right>}$ 
of the vacuum 
$\left<A_r\right> = \left<A_1A_2\cdots A_{\NC}\right>$ 
by 
\beq
{\cal W}^{\left<A_r\right>} (x^1,x^2) \equiv
\sum_{r=1}^{\NC} \left(m_{A_r}x^1 + n_{A_r}x^2\right) 
+ a^{\left<A_r\right>}.
\label{eq:vacuum_weight}
\eeq
Here we defined $a^{\left<A_r\right>}$ as a real part of 
$\log\det H_0^{\left<A_r\right>} 
= a^{\left<A_r\right>} + ib^{\left<A_r\right>}$,
where $H_0^{\left<A_r\right>}$ is $\NC \times \NC$ 
minor matrix whose elements are given by 
$\left(H_0^{\left<A_r\right>}\right)^s{}_t = \left(H_0\right)^s{}_{A_t}$.
Then $\Omega_0$ is given by 
\begin{equation}
\det \Omega_0
\equiv \det\left(\frac{1}{c}H_0 e^{2(M_1x^1+M_2x^2)} H_0^\dag \right)
=\log\left(
\sum_{\langle A_r \rangle}e^{2{\cal W}^{\langle A_r \rangle}} \right)
.
\label{eq:omega0_vacuum_weight}
\end{equation}
Since the solution of the master equation $\Omega$ is 
well-approximated by $\Omega_0$ near vacuum regions, 
we can use the weight of the vacuum ${\cal W}^{\langle A_r \rangle}$ 
to estimate the regions where the vacuum $\langle A_r \rangle$ 
is dominant. 
Therefore, position of the domain wall 
which divides two vacua $\left<A_r\right>$ and $\left<B_r\right>$ (all the other weights are 
much smaller than these two) can be estimated
from the condition of equating the weights of the vacua 
${\cal W}^{\left<A_r\right>} \simeq {\cal W}^{\left<B_r\right>}$:
\beq
\sum_{r=1}^{\NC}\left(m_{A_r}-m_{B_r}\right) x^1 +
\sum_{r=1}^{\NC}\left(n_{A_r}-n_{B_r}\right) x^2 +
a^{\left<A_r\right>} - a^{\left<B_r\right>} \simeq 0.
\label{eq:posi-ang}
\eeq
Hence the parameter 
$a^{\left<A_r\right>} - a^{\left<B_r\right>}$ in the moduli 
matrix 
determines the position of the domain wall. 
Furthermore, one can see the angle of the domain wall
is determined  by the mass difference between the two vacua.
Notice that the domain wall line (\ref{eq:posi-ang}) is 
perpendicular to the corresponding 
edge of the grid diagram, see Fig.\ref{fig:grid}. 
So the grid diagram gives us information of 
the angle of domain wall in addition to the tension. 
A junction point at which three of domain walls get 
together can also be estimated by the condition of equating 
the weights of these three related vacua as 
${\cal W}^{\left<A_r\right>} 
\simeq {\cal W}^{\left<B_r\right>} 
\simeq {\cal W}^{\left<C_r\right>}$.

\section{Effective Actions on Networks/Webs of Walls}\label{sec:effective}

\subsection{General Form of the Effective Action}\label{subsec:general}

Let us construct an effective theory on the world-volume of 
the networks/webs of walls. 
Zero modes on the background BPS solutions which are the 
lightest modes will play a main role in the effective theory 
while all the massive modes will be ignored in the following. 
Normalizable zero modes can be promoted to fields on the world-volume 
of the soliton background from mere parameters, with the 
assumption of weak dependence of the world-volume coordinates 
(slow move approximation {\it \`a la} Manton \cite{Manton:1981mp}). 
However, one should note that webs of walls contain 
nonnormalizable zero modes which cannot be promoted to 
effective fields and should be regarded as ``couplings'' 
of effective fields that are specified by boundary 
conditions\cite{Eto:2005cp,Eto:2005fm}. 
In the case of webs of walls, all the moduli parameters $\phi^i$ 
are contained in the moduli matrix $H_0(\phi^i)$. 
So we promote normalizable modes 
to fields on the world-volume coordinates $x^\mu~~(\mu=0,3)$
\begin{align}
H_0\bigl(\phi^i \bigr) \rightarrow H_0 \bigl(\phi^i(x^\mu)\bigr) \label{eq:prom}.
\end{align}
Now we introduce ``the slow-movement order parameter" $\lambda$, 
which is assumed to be much smaller than 
the other typical mass scales in the problem. 
There are two characteristic mass scales: 
one is mass difference $|\Delta m|$ of hypermultiplets, 
and the other is $g\sqrt{c}$ in front of the master equation. 
Therefore, we assume that
\begin{align}
\lambda \ll \text{min}(|\Delta m|,g\sqrt{c}).
\end{align}
The non-vanishing fields in the 1/4 BPS background have
contributions independent of $\lambda$, namely we assume that
\begin{align}
H^1=\mathcal{O}(1),
\quad W_\alpha = \mathcal O(1),
\quad \Sigma_\alpha=\mathcal{O}(1).
\end{align}
The derivatives of these fields with respect to the world-volume coordinates 
are assumed to be of order $\lambda$
expressing the weak dependence on the world-volume coordinates. 
The vanishing fields in the background can now have non-vanishing values, induced by the fluctuations 
of the moduli fields of order $\lambda$. Therefore, we assume that
\begin{align}
\partial_\mu=\mathcal{O}(\lambda),
\quad W_\mu=\mathcal{O}(\lambda),
\quad H^2=\mathcal{O}(\lambda).
\end{align}
Then the covariant derivative $\D_\mu = {\cal O}(\lambda)$ has consistent $\lambda$ dependence.

If we expand the full equations of motion of the Lagrangian 
(\ref{eq:lag}) in powers of $\lambda$, then we find that the 
$\mathcal{O}(1)$ equations 
are automatically satisfied due to the BPS equations.
The next leading order $\mathcal{O}(\lambda)$ equation 
is the equation for $W_\mu$ including the Gauss law 
\begin{align}
0=-\frac{2}{g^2}\mathcal D_\alpha F_{\alpha\mu}+\frac{2i}{g^2}
(\Sigma_\alpha \mathcal D_\mu \Sigma_\alpha
-\mathcal D_\mu \Sigma_\alpha \Sigma_\alpha)
+i(H^1\mathcal D_\mu H^{1\dag}-\mathcal D_\mu H^1 H^{1\dag}). 
\label{eq:lambda}
\end{align}
In order to obtain the effective action of order 
$\lambda^2$ on webs of walls, 
we have to solve this equation and determine the 
configuration of $W_\mu$.

Eq.~(\ref{eq:lambda}) contains the derivatives of the world-volume 
coordinates $x^\mu ~(\mu=0,3)$ on wall webs. 
As a consequence of Eq.(\ref{eq:prom}), 
the moduli matrix $H_0(\phi^i)$ depends on the world-volume 
coordinates $x^\mu$ through the moduli fields $\phi^i(x^\mu)$. 
Note that for the fields which depend on $x^\mu$ only 
through the moduli fields, the derivatives with respect to 
the world-volume coordinates satisfy 
\begin{align}
\partial_\mu=\delta_\mu+\delta_\mu^\dag,
\end{align}
where we have defined the differential operators 
$\delta_\mu$ and $\delta_\mu^\dagger$ by
\begin{align}
\delta_\mu \equiv 
\sum_i \partial_\mu \phi^i \frac{\delta}{\delta \phi^i},
\quad
\delta_\mu^\dag 
\equiv \sum_i \partial_\mu \phi^{i\ast} 
\frac{\delta}{\delta \phi^{i\ast}},
\end{align}
respectively. 
Using these operators, 
the $\mathcal O(\lambda)$ equation of motion Eq.(\ref{eq:lambda}) 
can be solved, to yield 
\begin{align}
W_\mu=i(\delta_\mu S^\dag S^{\dag-1}-S^{-1}\delta_\mu^\dag S).
\end{align}

The remaining work is to substitute these solutions into 
the fundamental Lagrangian (\ref{eq:lag}) and to 
integrate over the codimensional coordinates $x^1$ and $x^2$. 
Since we are interested in the leading 
nontrivial part in powers of $\lambda$, we retain the terms up to $\mathcal O(\lambda^2)$. 
We ignore a total derivative term which does not 
contribute to the effective Lagrangian, and obtain 
the effective Lagrangian of order ${\cal O}(\lambda^2)$ 
as 
\begin{eqnarray}
\mathcal{L}^{eff} 
&=&\delta_\mu \delta^{\mu \dag} K(\phi,\phi^*)
 =K_{ij^*}(\phi,\phi^*)\partial^\mu \phi^i 
\partial_\mu \phi^{j*}, \label{eq:general} \\
K(\phi,\phi^*)
&=& 
K_w (\phi,\phi^*) + K_g (\phi,\phi^*) 
\label{eq:kah}
\\
K_w(\phi,\phi^*) 
&\equiv& \int d^2x\, 
c\,\text{log det}\Omega, 
 \label{eq:kah_w}
\\
K_g(\phi,\phi^*) 
&\equiv& \int d^2x\,
\frac{1}{2g^2}\text{Tr}(\Omega^{-1}\partial_\alpha
\Omega)^2 
. \label{eq:kah_g}
\end{eqnarray}
From the above expression, we can see that the metric 
$K_{ij^*}$ on the moduli space 
is a K\"ahler metric whose K\"ahler potential 
is given by Eq.~(\ref{eq:kah}). 
We observe that the above formula is valid for the 
effective Lagrangian on the domain wall, if we restrict 
ourselves to single codimension, for instance $\alpha=1$ 
and integrate over $x^1$ alone when the wall is 
perpendicular to the $x^1$ axis. 
Note that the integral in Eqs.~(\ref{eq:kah_w}) 
and (\ref{eq:kah_g}) for the 
K\"ahler potential has an infrared divergence 
($x^1, x^2 \to \infty$). 
This comes from the fact that we have only constituent 
walls asymptotically ($x^1, x^2 \to \infty$). 
Since we have not  promoted the nonnormalizable zero modes 
associated to the constituent walls 
and have fixed them by boundary conditions,   
these apparent contributions from asymptotic infinity 
to the K\"ahler potential can actually be eliminated by a 
K\"ahler transformation. 
Namely the infrared divergence can be removed 
by subtracting, from the integrand of Eqs.~(\ref{eq:kah_w}) 
and (\ref{eq:kah_g}), 
some functions which are holomorphic or anti-holomorphic 
with respect to the effective fields.

We eventually got a K\"ahler geometry as a target space 
of a nonlinear sigma model on the moduli space: 
we did not expect K\"ahler geometry for effective action because 
it is $d=2$, ${\cal N}=(2,0)$ supersymmetric nonlinear sigma model
\cite{Eto:2005cp,Eto:2005sw} 
whose geometry does not have to be K\"ahler in general 
\cite{Hull:1985jv,Witten:2005px}.\footnote{
If we have an anti-symmetric tensor field $b_{ij}$ in the 
effective theory, the geometry is no longer K\"ahler but 
becomes K\"ahler (Hermitian) 
with torsion \cite{Hull:1985jv,Witten:2005px}. 
This may happen if we add a theta term in the original Lagrangian. 
}


\subsection{Single Triangle Loop}\label{subsec:1-loop}

Let us first consider the effective action on a single 
triangle loop of walls. Here we will briefly review
the construction of single triangle loop. In order to obtain 
a single triangle loop, we need at least four vacua. 
Here we consider the model with $\NC=1$, $\NF=4$ and masses 
\begin{align}
M \equiv M_1+iM_2=\text{diag}(m_1+in_1, m_2+in_2, m_3+in_3, m_4+in_4).
\end{align}
The model with $\NC=3$, $\NF=4$ will be discussed 
in Sec.~\ref{subsec:trononabelian}. 
In the $A$-th 
vacuum $\langle A \rangle$, 
the adjoint scalar takes the following vacuum expectation value:
\begin{align}
\langle \Sigma \rangle_{\langle A \rangle}
\equiv \langle \Sigma_1+i\Sigma_2 \rangle_{\langle A \rangle}
=m_A+in_A.
\label{eq:sigma}
\end{align}
The moduli matrix which determines the configuration of walls is now 
a four-component row vector 
\begin{align}
H_0=\sqrt{c}(e^{a_1+ib_1},e^{a_2+ib_2},e^{a_3+ib_3},e^{a_4+ib_4}).
\end{align}
Now $\Omega_0$ is given by 
\begin{eqnarray}
 \Omega_0 \equiv \frac{1}{c}H_0 e^{2(M_1x^1+M_2x^2)} H_0^\dagger 
 = \sum_{A=1}^{\NF} e^{2\mathcal{W}^{\langle A \rangle}} .
\end{eqnarray} 
Here we define the weight of vacuum $\langle A \rangle$ as
\begin{align}
 e^{2\mathcal{W}^{\langle A \rangle}} 
 \equiv e^{2(m_Ax^1+n_Ax^2+a_A)}.
\end{align}

\begin{figure}[h]
\begin{center}
\includegraphics[height=4cm]{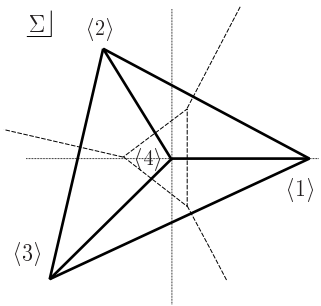}
\caption{Grid diagram and web diagram}
\label{fig:cp3_sc_loop}
\end{center}
\end{figure}

In order to obtain a single triangle loop, we have to set 
mass parameters appropriately. 
Recall that the web diagram in the configuration space is dual 
of the grid diagram in the complex $\Sigma$ plane \cite{Eto:2005cp}.
Therefore, a single triangle loop appears when the grid 
diagram 
consists of a triangle and 
one vacuum 
point is located inside the triangle 
(See Fig.(\ref{fig:cp3_sc_loop})).
For simplicity, we set the fourth complex mass parameter 
to zero and impose the following conditions:
\begin{align}
       \vec m_1 \times \vec m_2 >0,
 \quad \vec m_2 \times \vec m_3 >0, 
 \quad \vec m_3 \times \vec m_1 >0,
 \label{eq:mass}
\end{align}
where $\vec m_i \times \vec m_j \equiv m_in_j-m_jn_i$.

Before constructing the effective action of the wall web, 
we have to know which mode is normalizable and which is not. 
In the present case, 
there exist three walls as external legs 
(we call them {\it external walls}): 
the wall 
interpolating $\langle 1 \rangle$ and $\langle 2 \rangle$, 
the one interpolating  $\langle 2 \rangle$ and $\langle 3 \rangle$ and 
the one interpolating  $\langle 3 \rangle$ and $\langle 1 \rangle$. 
These three walls have semi-infinite length. 
There also exist three {\it internal walls} as segments: 
the wall interpolating $\langle 1 \rangle$ and $\langle 4 \rangle$, 
the one interpolating $\langle 2 \rangle$ and $\langle 4 \rangle$ and 
the one interpolating 
$\langle 3 \rangle$ and $\langle 4 \rangle$.  
They constitute the single triangle loop.   
Zero modes which are related to the positions of the 
semi-infinite external walls, that is, zero modes which 
change the boundary condition at infinity are 
non-normalizable modes in the effective theory 
of the wall web. 
We should fix such non-normalizable modes.
Therefore, the zero mode which is related to the size of 
the loop is the only possible normalizable mode.

\begin{figure}[h]
\begin{center}
\includegraphics[width=4cm]{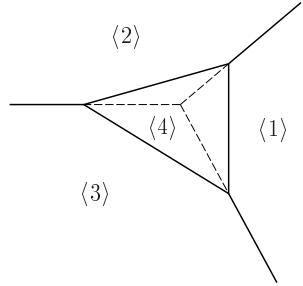}
\caption{Configuration of the wall loop}
\label{fig:loop2}
\end{center}
\end{figure}

In order to fix the three external walls, 
we fix three complex moduli parameters to 1 
with leaving single complex parameter 
$\phi=e^{r+i\theta}$ 
which corresponds to the (dimensionless) size $r$ of the 
loop and the associated phase moduli $\theta$:
\begin{align}
H_0=\sqrt{c}(1,1,1,\phi)~~~~\text{with}~~\phi=e^{r+i\theta}.
\label{eq:A_triangle_moduli}
\end{align}
We promote this moduli parameter $\phi$ to a field on 
the world volume and construct the effective action for that.
The configuration of the wall loop is given in Fig.\ref{fig:loop2}.

Next we have to solve the master equation (\ref{eq:master}) 
to explicitly construct the effective action. 
In order to do that, we take the strong coupling limit 
$g^2 \rightarrow \infty$, in which the master equation 
can be solved algebraically as $\Omega=\Omega_0$. 
In the present case, $\Omega_0$ is given by
\begin{align}
\Omega_0=e^{2\vec m_1 \cdot \vec x}
+e^{2\vec m_2 \cdot \vec x}+e^{2\vec m_3 \cdot \vec x}+|\phi|^2.
\label{eq:omega0}
\end{align}
Substituting this solution to Eq.(\ref{eq:kah}) and 
neglecting the second term ($g^2\to \infty$), 
we obtain the following K\"ahler potential 
\begin{align}
K = c \int d^2x \log\Omega_0 
  = c \int d^2x \log(e^{2\vec m_1 \cdot \vec x} 
   + e^{2\vec m_2 \cdot \vec x}
+e^{2\vec m_3 \cdot \vec x}+|\phi|^2).
\label{eq:kah-g-inf}
\end{align}
Though this is apparently divergent, we can redefine the 
K\"ahler potential using the K\"ahler transformation to 
remove the moduli-independent divergence:
\begin{align}
K_w &\equiv c\,\int d^2x 
\left[\,\log\Omega_0-\log\,\tilde{\Omega}_0\,\right] 
=c\,\int d^2x \,\log\left(
                  1+\frac{|\phi|^2}{\tilde{\Omega}_0}\right), 
  \label{eq:modkah} \\
\tilde \Omega_0 &\equiv
   e^{2\vec m_1\cdot \vec x} + e^{2\vec m_2\cdot \vec x} 
 + e^{2\vec m_3\cdot \vec x}.
 \label{eq:omegatil}
\end{align}

We will show in the next section that this K\"ahler potential 
can be written as a sum of hypergeometric functions. 
We will obtain asymptotic metric 
in Sec.~\ref{sec:tropical} by using another simple approach,
which is valid for sufficiently large loop configuration $\phi \gg 1$. 
We will also calculate the asymptotic value of the second term 
$K_g$ in Eq.~(\ref{eq:kah}) in Sec.~\ref{sec:tropical} 
without using explicit solutions.
The meaning of the asymptotic metric will be clarified 
in Sec.~\ref{sec:understand}.

\subsection{K\"ahler Potential in terms of Hypergeometric Functions}\label{subsec:hypergeometric}

Before trying to compute the K\"ahler potential, we first 
introduce the following notations:
\begin{align}
&\Delta_{[123]}
=\vec m_1\times \vec m_2 +\vec m_2 \times \vec m_3 
+\vec m_3 \times \vec m_1, \\
&\alpha_1\equiv \frac{(\vec m_2\times \vec m_3)}{\Delta_{[123]}},\quad
\alpha_2\equiv \frac{(\vec m_3\times \vec m_1)}{\Delta_{[123]}},\quad
\alpha_3\equiv \frac{(\vec m_1\times \vec m_2)}{\Delta_{[123]}}.
\end{align}
Since the quantity $\tilde \Omega_0$ 
in Eq.(\ref{eq:omegatil}) 
has the following minimum value 
\begin{eqnarray}
\log \tilde \Omega_{\rm min}=-\sum_{i=1}^3\alpha_i\log\alpha_i 
\end{eqnarray}
as a function of $x^1, x^2$, 
we can expand the integrand in powers of 
$\frac{|\phi|^{2}}{\tilde \Omega_0}$ 
as long as $|\phi|^2 
\le \exp \left( -\sum \alpha_i\log\alpha_i \right) $ 
\begin{eqnarray}
K_w=c\,\sum_{n=1}^\infty \frac{(-1)^{n+1}}{n}
\int d^2x \frac{|\phi|^{2n}}{\tilde \Omega^n_0}.
\end{eqnarray}
Let us define the new variables of integration by 
\beq
s_1 = (\vec m_1-\vec m_3) \cdot \vec x,\hs{5} s_2 
= (\vec m_2-\vec m_3) \cdot \vec x.
\eeq
Then, for each $n$, we can integrate this in the following way:
\begin{eqnarray}
 \int^\infty_{-\infty} \frac{dx^1dx^2}{\tilde \Omega^n_0}
&=&\frac{1}{\Delta_{[123]}}\int^\infty_{-\infty} ds_1ds_2
\frac{e^{2n\alpha_1 s_1}e^{2n \alpha_2 s_2}}
{(e^{2s_1}+e^{2s_2}+1)^n}\notag \\
&=&\frac{1}{4\Delta_{[123]}}\int^{\infty}_0 dudv
\frac{u^{n \alpha_1-1}v^{n\alpha_2-1}}{\left(u+v+1\right)^n} \qquad (u=e^{2s_1}, v=e^{2s_2})\notag \\
&=&\frac{1}{4\Delta_{[123]}\Gamma(n)}\int^{\infty}_0 dudvds\,
u^{n \alpha_1-1}v^{n\alpha_2-1}s^{n-1}e^{-\left(u+v+1\right)s}\notag \\
&=&\frac{1}{4\Delta_{[123]}}\frac{\Gamma(\alpha_1n)
\Gamma(\alpha_2n)\Gamma(\alpha_3n)}{\Gamma(n)}.
\end{eqnarray}
Therefore we obtain the final form of the K\"ahler potential $K_w$ as 
\begin{align}
K_w=
\frac{c}{4\Delta_{[123]}}
\sum_{n=1}^\infty \frac{(-1)^{n+1}}{n}\frac{\Gamma(\alpha_1n)
\Gamma(\alpha_2n)\Gamma(\alpha_3n)}{\Gamma(n)}|\phi|^{2n}.
\label{eq:kahlerw}
\end{align}
We see that this series is convergent for 
$|\phi|^2 \le \exp \left( -\sum \alpha_i\log\alpha_i \right)$ 
and defines an analytic function of $|\phi|^{2}$ with a 
possible pole on the negative real axis. 
Therefore we know that there exists well-defined smooth 
function even for 
$|\phi|^2 \ge \exp \left( -\sum \alpha_i\log\alpha_i \right)$. 
Moreover, this can be written as a sum of hypergeometric functions 
in cases where ${\alpha_i}$ 
are rational numbers. 
We can find out the behavior of this function outside 
the convergence region of the series 
by using the analytic continuation. 
In Appendix A, we give 
the detail of the analytic continuation to study the 
asymptotic behavior of a more general power series 
expression (\ref{eq:Gpfunction}) including $K_w$ as a 
special case, and the expression 
in terms of hypergeometric functions. 
In the next section, we will show another simple approach 
to know the asymptotic metric for $|\phi|^2 \gg 1$. 

By using the expression Eq.~(\ref{eq:kahlerw}), we can 
calculate the metric for 
$|\phi|^2 \le \exp \left( -\sum \alpha_i\log\alpha_i \right)$ as
\beq
K_{\phi \bar \phi} = \frac{\p^2 K_w}{\p \phi \p \bar \phi} 
= \frac{c}{4\Delta_{[123]}}
\sum_{n=1}^\infty (-1)^{n+1} n \frac{\Gamma(\alpha_1n)
\Gamma(\alpha_2n)\Gamma(\alpha_3n)}{\Gamma(n)}|\phi|^{2(n-1)}.
\label{eq:metric}
\eeq
Since the loop of walls shrinks completely at $\phi=0$, 
one might think that the moduli space is singular at $\phi=0$. 
However, we can 
exactly calculate the scalar curvature from 
Eq.~(\ref{eq:metric}) at $\phi=0$ 
and find that it is finite:
\beq
R 
= \frac{16\Delta_{[123]}}{c} \frac{\Gamma(2 \alpha_1)
\Gamma(2 \alpha_2)\Gamma(2 \alpha_3)}{\left(\Gamma(\alpha_1)
\Gamma(\alpha_2)\Gamma(\alpha_3)\right)^2} + \mathcal O(|\phi|^2).
\eeq
Therefore, the moduli space of the single loop is 
non-singular at $\phi=0$.
\begin{figure}[h]
\begin{center}
\includegraphics[width=60mm]{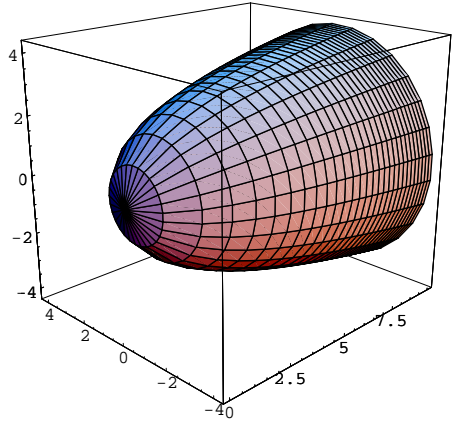}
\end{center}
\caption{The moduli space of single triangle loop embedded 
in $\mathbf R^3$: 
The moduli space has a $U(1)$ isometry which corresponds 
to the direction of the phase modulus. 
The other direction 
can be regarded as the direction of size modulus of the loop. 
The tip of the moduli space corresponds to the point 
$\phi=0$ where the loop shrinks completely.}
\label{fig:metric}
\end{figure}
Fig.~\ref{fig:metric} illustrates the moduli space of the single 
loop embedded into the three dimensional Euclidean space.

\section{Large Size Behavior: Tropical Limit}\label{sec:tropical}

\subsection{Tropical Limit 
}
\label{subsec:tropical}

To evaluate the asymptotic metric of the moduli space at 
large loop size, we note that $\log {\rm det}\Omega$ is well 
approximated by $\log {\rm det}\Omega_0$ apart from tiny finite 
regions near walls and junctions. 
By excluding these finite regions near walls and junctions, 
space is divided into a number of regions corresponding to 
various vacua. 
In each region, 
only 
the weight of a single vacuum is 
dominant in $\log {\rm det}\Omega_0$
\begin{equation}
\log\det \Omega_0
=\log\left(
\sum_{\langle A_r \rangle}e^{2{\cal W}^{\langle A_r \rangle}} \right)
\to 
 \underset{\langle A \rangle}
{\rm max} 
\,\left( 2 {\cal W}^{\langle A \rangle} \right)
,
\end{equation}
where the weight of the vacuum ${\cal W}^{\langle A \rangle}$ 
is defined in Eqs.(\ref{eq:vacuum_weight}) and 
(\ref{eq:omega0_vacuum_weight}). 
We call this procedure ``tropical limit". 
It will turn out that the tropical limit exactly extracts 
the leading contribution from the K\"ahler potential and 
the correction terms are strongly suppressed in 
sufficiently large loop configuration $|\phi|^2 \gg 1$. 
Moreover, the tropical limit is applicable irrespective of 
the value of gauge coupling $g$. 

As a concrete example, let us take the Abelian gauge theory 
with four flavors. 
The configuration of single triangle loop has three 
external walls and three internal walls, which divide 
the $x^1$-$x^2$ plane into four regions, 
$\langle 1 \rangle$, $\langle 2 \rangle$, 
$\langle 3 \rangle$ and $\langle 4 \rangle$ (See Fig.~\ref{fig:loop2}). 
In each region $\langle A \rangle,\,(A=1,2,3,4)$, 
the corresponding weight $e^{2{\cal W}^{\langle A \rangle}}$ is 
largest among four weights. 
In the present section, we will consider only the largest 
weight in each region and simply drop the other weights 
to roughly evaluate the K\"ahler potential. 
Namely, we consider the limit in which the logarithmic 
function becomes 
\beq
\log \Omega_0 = \log \left( e^{2{\cal W}^{\langle 1 \rangle}} + e^{2{\cal W}^{\langle 2 \rangle}} + e^{2{\cal W}^{\langle 3 \rangle}} + e^{2{\cal W}^{\langle 4 \rangle}} \right) \rightarrow \underset{1 \leq A \leq 4}{\rm max} \,\left( 2 {\cal W}^{\langle A \rangle} \right)
\eeq

\subsection{Tropical Limit in 1/2 BPS Parallel Walls}
\label{subsec:parallel}

\begin{figure}[ht]
\begin{center}
\begin{tabular}{cc}
\includegraphics[width=70mm]{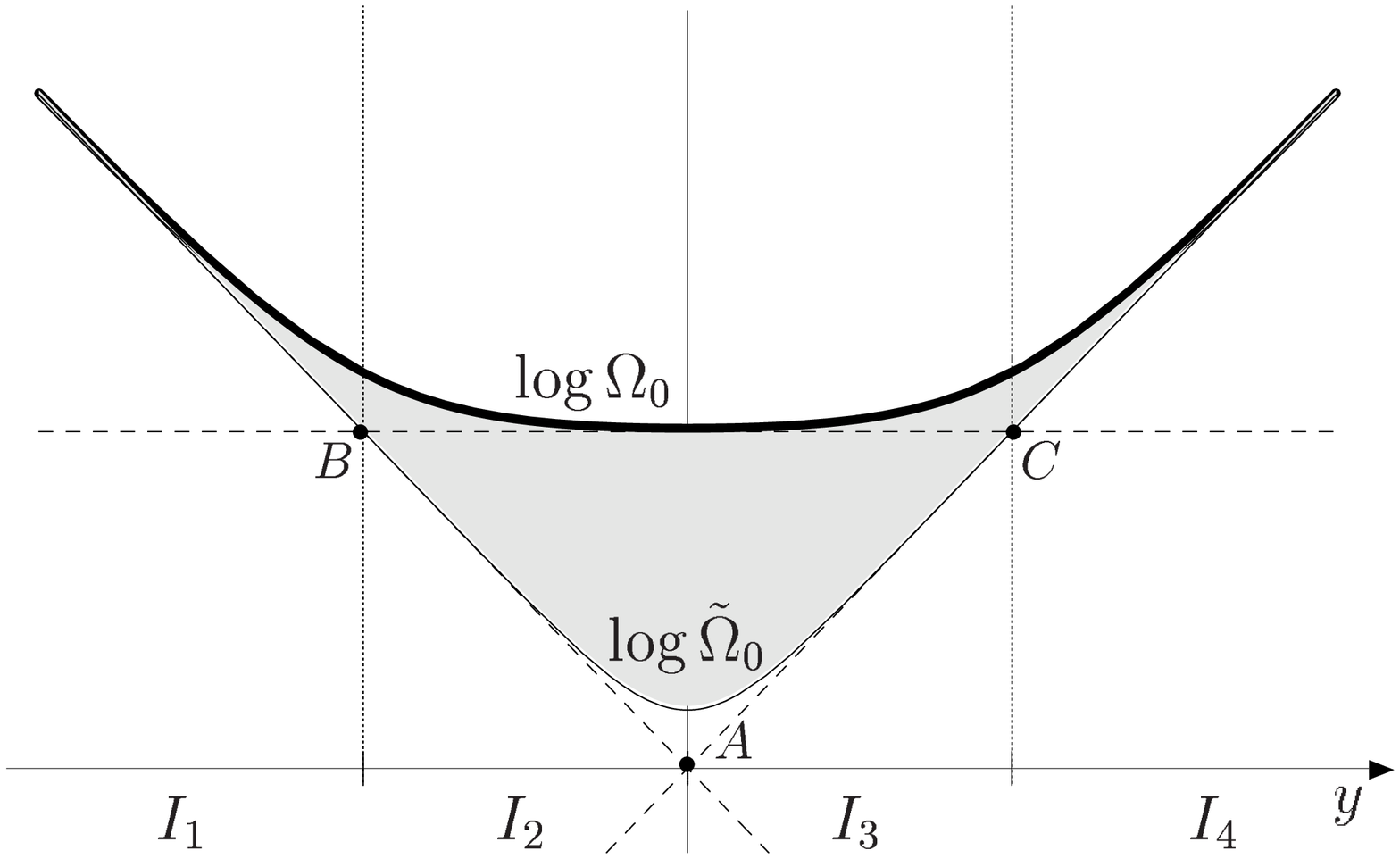} ~~~~~ &
\includegraphics[width=70mm]{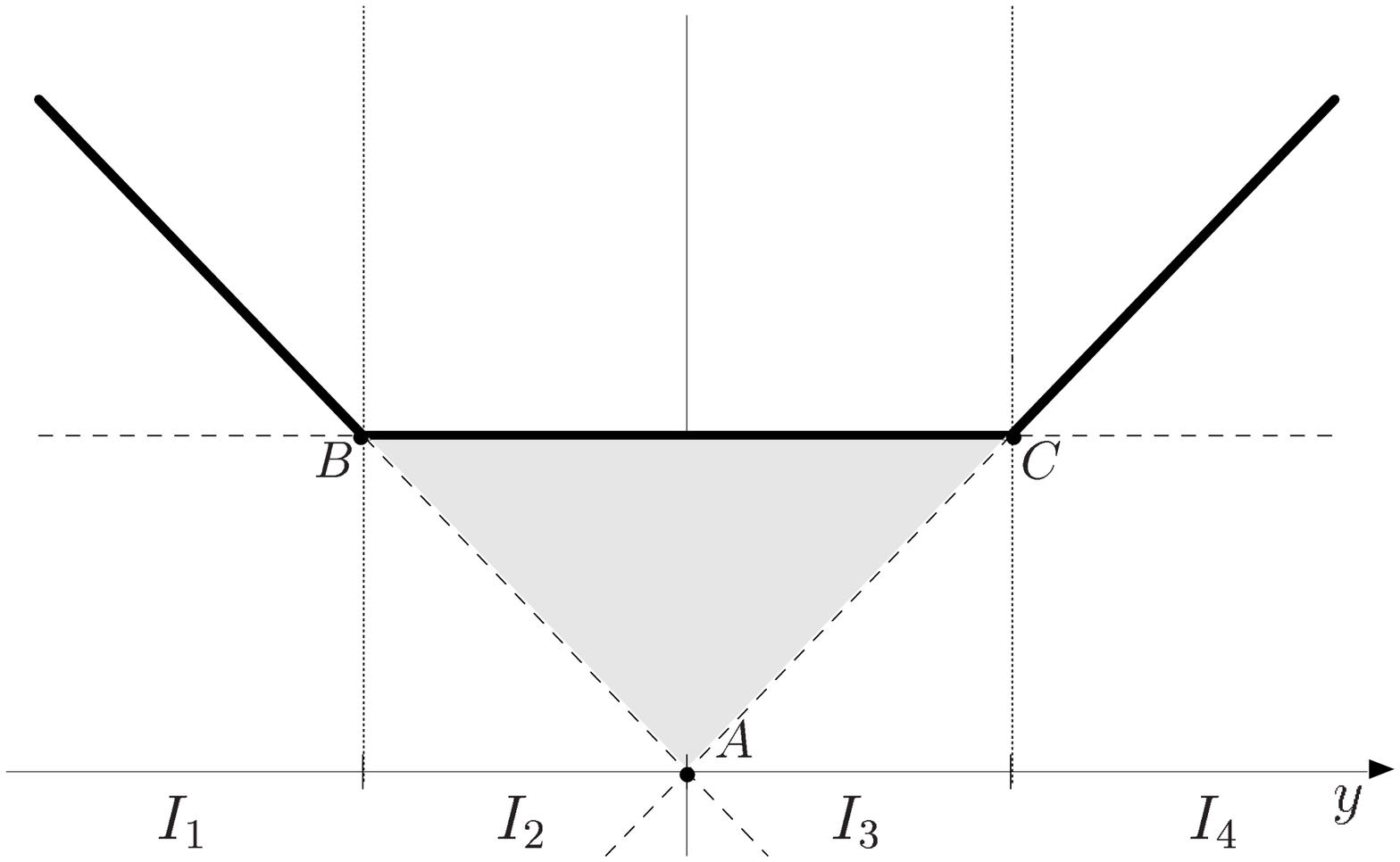} ~~~~~\\
(a) Plot of $\log \Omega_0$ and $\log \tilde \Omega_0$&
(b) Tropical limit of $\log \Omega_0$
\end{tabular}
\end{center}
\caption{ 
The K\"ahler potential is given by the area of the shaded region in (a). The area of the triangle $ABC$ (first term in Eq.~(\ref{eq:Kahler})) gives the leading contribution to the K\"ahler potential for large $r$.}
\label{fig:para}
\end{figure}

Before considering the case of large $|\phi|^2$ of the 
loop of the walls, 
let us consider 
simpler example of two parallel walls and evaluate 
concretely an effective K\"ahler potential for 
a relative distance (and a relative phase) between two 
walls. 
This computation should serve as a warm-up exercise for 
the more complicated case of wall loops. 
The model for $\NC=1$ and $\NF=3$ with non-degenerate real
masses $m_A (A=1,2,3)$
have three different vacua characterized by
$\Sigma_1+i\Sigma_2=m_1,m_2,m_3\in {\bf R}$
and allows a BPS solution of double walls with two complex moduli
parameters. 
We can take one 
of the two 
to be the center of mass position and the overall phase 
which are Nambu-Goldstone modes and normalizable in this case 
of parallel walls. 
Therefore we promote it to free particles 
in the effective action. 
Let us now concentrate to 
the other complex moduli $\phi=e^{r+i\theta}$. 
A real part $r$ can be interpreted 
as the relative distance if $|\phi|>1$, 
and it's imaginary part $\theta$ gives the relative phase 
between two walls.  
A K\"ahler potential of an effective action 
for a chiral field $\phi$ in the strong coupling
limit is obtained by performing the following 
integral~\cite{Eto:2006uw}, 
\begin{eqnarray}
 K_{\rm wall}=c\,\int_{-\infty}^{\infty}dy\left\{\log\left(
e^{2m\alpha_1y}+e^{-2m\alpha_2y}+|\phi|^2\right)-\log\left(
e^{2m\alpha_1y}+e^{-2m\alpha_2y}\right)\right\}
\end{eqnarray}
where we denoted the mass parameters as 
$(m_1,m_2,m_3)=(m \alpha_1,\,0,\, -m\alpha_2)$ 
with $m\equiv m_1-m_3$, $\alpha_1+\alpha_2=1$ and 
$0<\alpha_1<1$ without 
loss of generality.\footnote{
By shifting the adjoint scalar $\Sigma_\alpha$, one can 
always choose one of the masses to vanish. 
}
To perform 
the integration for large $r$, let us decompose the
integral into four integrals with respect to 
intervals $I_i, (i=1,\cdots,4)$ 
\begin{eqnarray}
&& I_1\equiv\{y|e_1\ge e_0\ge e_2\}
=\big\{y\big |my\ge \frac{r}{\alpha_1}\big\},\nn
&& I_2\equiv\{y|e_0>e_1\ge e_2\}
=\big\{y\big |\frac{r}{\alpha_1}>my\ge 0\big\},\nn
&& I_3\equiv\{y|e_0\ge e_2>e_1\}
=\big\{y\big |0>my\ge -\frac{r}{\alpha_2}\big\},\nn
&&I_4\equiv\{y|e_2>e_0>e_1\}
=\big\{y\big |my< -\frac{r}{\alpha_2}\big\},
\end{eqnarray}
with abbreviations 
$e_1=e^{2m\alpha_1y}, e_2=e^{-2m\alpha_2y}, e_0=e^{2r}$. 
Here, note that 
 $\frac{r}{\alpha_1},-\frac{r}{\alpha_2}$ are (dimensionless) 
positions of walls and the center of mass position is chosen 
to vanish. The various informations are illustrated in Fig.~\ref{fig:para}.
In the interval $I_1$, it is convenient to perform the integral 
by decomposing 
the integrand into  
\begin{eqnarray}
&& \log(e_1+e_2+e_0)-\log(e_1+e_2)\nn
&=&\log\left(1+\frac{e_0}{e_1}\right)
-\log\left(1+\frac{e_2}{e_1}\right)
+\log\left(1+\frac{e_2}{e_1+e_0}\right),
\label{eq:wall-integral-1}
\end{eqnarray}
so that logarithmic function can be expanded to infinite series as
$\log(1+x)=\sum_{n=1}^\infty\frac{(-1)^{n-1}}{n}x^n$.
Note that the term $\log e_1$ which gives divergent 
integral cancels out. 
In the interval $I_2$, we also take a decomposition of the 
integrand as
\begin{eqnarray}
\left(\log e_0-\log e_1\right)+\log\left(1+\frac{e_1}{e_0}\right)
-\log\left(1+\frac{e_2}{e_1}\right)
+\log\left(1+\frac{e_2}{e_1+e_0}\right),
\label{eq:wall-integral-2}
\end{eqnarray}
The first term gives leading terms of the integral as
 \begin{eqnarray}
  \int_{I_2}dy\left(\log e_0-\log e_1\right) = \int_{\frac{r}{m\alpha_1}}^0 dy\, \left(2r - 2 m \alpha_1 y \right) = \frac{1}{m\alpha_1}r^2,
 \end{eqnarray}
and next leading terms are calculated as
\begin{eqnarray}
 \int_{I_1}dy\log\left(1+\frac{e_0}{e_1}\right)
&=&\sum_{n=1}^\infty\frac{(-1)^{n-1}}{n}e^{2 n r}
\int_{\frac{r}{\alpha_1}}^{\infty}dy\,e^{-2 n \alpha_1y}
=\frac{1}{2m\alpha_1} \frac{\pi^2}{12},
\qquad \quad
\label{eq:nextleading1}
\\
 \int_{I_2}dy\log\left(1+\frac{e_1}{e_0}\right)
&=&\frac1{2m\alpha_1}
\left(\frac{\pi^2}{12}
-\sum_{n=1}^\infty\frac{(-1)^{n-1}}{n^2}e^{-2n r}\right),
\label{eq:nextleading2}
\\
-\int_{I_1+I_2}dy\log\left(1+\frac{e_2}{e_1}\right)
&=&-\frac{1}{2m}\frac{\pi^2}{12},
\label{eq:nextleading3}
\end{eqnarray} 
where we used an identity
$\sum_{n=1}\frac{(-1)^{n-1}}{n^2}=\frac12\zeta(2)=\frac{\pi^2}{12}$.
We also find the 
integrals for 
the intervals $I_3$ and $I_4$ give the same 
contributions as $I_2$ and $I_1$, respectively, 
except for exchanging the parameters $\alpha_1$ and $\alpha_2$. 
There remain the last terms $\log(1+\frac{e_2}{e_1+e_0})$ 
in Eqs.(\ref{eq:wall-integral-1}) and (\ref{eq:wall-integral-2}). 
They should be integrated over $y$ in the region 
$I_1$ and $I_2$, respectively, and are evaluated in 
appendix \ref{appendix:correction}.  
The result can be combined with the last term (sum over 
$n$ of $e^{-2nr}$ terms) 
in Eq.(\ref{eq:nextleading2}) 
to give only 
contributions of order of $e^{-\frac{2}{\alpha_1}r}$. 
These correction terms are guaranteed to be finite by 
inequalities 
\begin{eqnarray}
 0<\int_{I_1,I_2}dy \log\left(1+\frac{e_2}{e_1+e_0}\right)
<\int_{I_1,I_2}dy \log\left(1+\frac{e_2}{e_1}\right)={\rm constant}.
\end{eqnarray}
Thus we obtain 
\begin{eqnarray}
 K_{\rm wall}=\frac{c}{\alpha_1\alpha_2m} r^2+
\frac{c}{2m}\left(\frac{1}{\alpha_1}+\frac{1}{\alpha_2}-1\right)
\frac{\pi^2}6+{\cal O}(e^{-\frac{2}{\alpha_{1}}r},
e^{-\frac{2}{\alpha_2}r}).
\label{eq:Kahler}
\end{eqnarray}
It is remarkable that there are no terms proportional to 
the length $r$ of the relative distance of walls in the 
K\"ahler potential. 
This is because the deviation from the tropical limit is 
localized in finite regions around points $A, B, C$ and 
does not increase as the distance $r$ between $B$ and $C$ 
increases. 
This feature holds irrespective of values of 
gauge coupling $g$. 
Therefore our approximation with the tropical limit 
as the leading term is applicable for any finite values of gauge coupling $g$.
One should note that the constant contribution (the second 
term) in the right-hand side of Eq.(\ref{eq:Kahler}) 
is a nontrivial physical constant, even though the term 
gives no contribution locally to the K\"ahler metric.  
For instance, if we fix the K\"ahler potential at $\phi=0$ 
($r=-\infty$) to vanish, we can no longer eliminate the 
constant term by K\"ahler transformations.

\subsection{Tropical Limit in Single Triangle Loop}
\label{subsec:troloop}

Our starting point is Eq.(\ref{eq:modkah}) 
where the moduli-independent divergence has already been 
removed. 
As we mentioned before, we divide the integral region to four parts, $\langle 1 \rangle, \langle 2 \rangle,
\langle 3 \rangle$ and $\langle 4 \rangle$ and pick up only the largest weight in each region. 
In integrating the second term, we also divide the integral 
region to three parts, and pick up only 
single weight which is largest in each region.

\begin{figure}[h]
\begin{center}
\includegraphics[width=6cm]{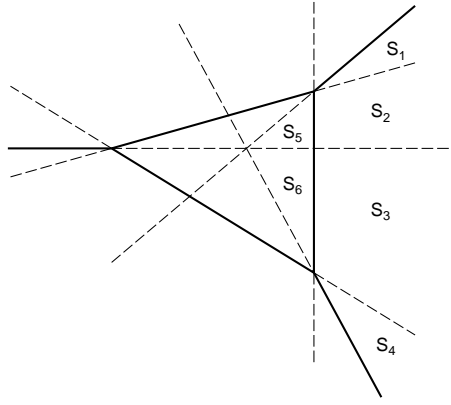}
\caption {Division of the integral region}
\label{fig:loop3}
\end{center}
\end{figure}

To illustrate what we are doing clearly, let us show 
the division of the integration region in 
Fig.\ref{fig:loop3}. 
If we denote each weight $e^{2{\cal W}^{\langle A \rangle}}$ 
as $e_A$, the order of four weights is  
$e_1 \ge e_2 \ge e_4 \ge e_3$ in the region $S_1$. 
The integrand in Eq.(\ref{eq:modkah}) can be rewritten as
\begin{align}
\log&(e_1+e_2+e_3+e_4)-\log(e_1+e_2+e_3)\notag \\
&= \log e_1+\log\left(1+\frac{e_2}{e_1}\right) 
 + \log \left(1+\frac{e_4}{e_1+e_2}\right)
 + \log \left(1+\frac{e_3}{e_1+e_2+e_4}\right) \notag \\
&~~- \log e_1- \log\left(1+\frac{e_2}{e_1}\right) 
   - \log \left(1+\frac{e_3}{e_1+e_2}\right)\notag \\
& = \log \left(1+\frac{e_4}{e_1+e_2}\right) 
  - \log \left(1+\frac{e_3}{e_1+e_2}\right) 
  + \log \left(1+\frac{e_3}{e_1+e_2+e_4}\right)
\end{align}
in the region $S_1$.
The leading contribution, $\log e_1$, cancels here. 
Similar cancellation occurs in 
$S_2, S_3$ and $S_4$. 
However, in the region $S_5$, the order of four weights 
is $e_4 \ge e_1 \ge e_2 \ge e_3$ and 
the integrand can be rewritten as 
\begin{align}
 \log&(e_1+e_2+e_3+e_4)-\log(e_1+e_2+e_3)\notag \\
&=\log e_4 + \log\left(1+\frac{e_1}{e_4} \right) 
 + \log \left(1+\frac{e_2}{e_4+e_1}\right)
+\log \left(1+\frac{e_3}{e_4+e_1+e_2}\right)\notag \\
&~~- \log e_1 - \log \left(1+\frac{e_2}{e_1}\right) 
   - \log \left(1+\frac{e_3}{e_1+e_2}\right).
\end{align}
The tropical limit yields non-zero value $\log e_4-\log e_1$ 
for the leading contribution in the region $S_5$. 
Similarly, in the region $S_6$, the leading contribution 
is $\log e_4-\log e_1$.
Contributions from the other regions 
can be obtained by rotating the label $i (i=1,2,3)$ cyclically.  

\begin{figure}[ht]
\begin{center}
\begin{tabular}{ccc}
\includegraphics[width=5cm]{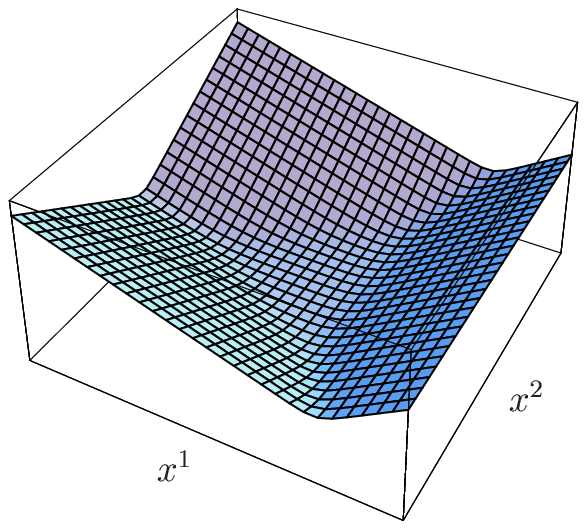}&
\includegraphics[width=5cm]{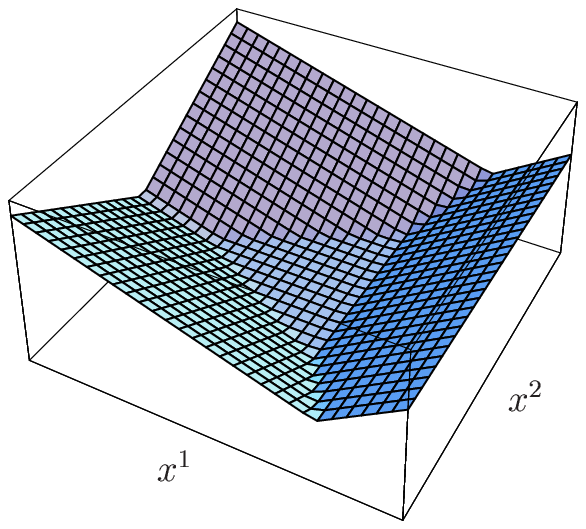}&
\includegraphics[width=5cm]{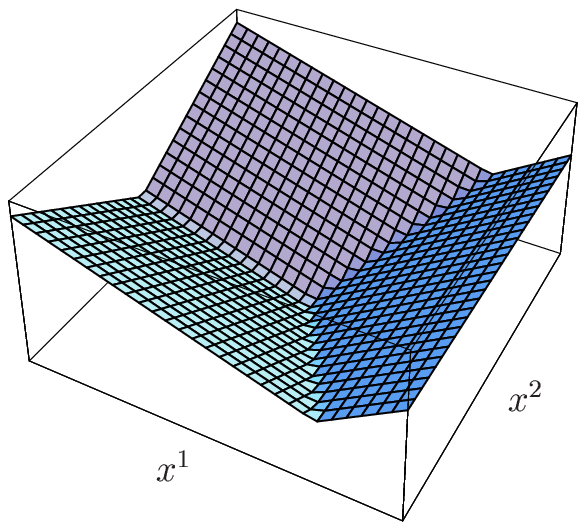}\\
(a)&(b)&(c)
\end{tabular}
\caption{(a)\,plot of $\log \Omega_0,$\quad(b)\,
tropical limit of $\log \Omega_0$,\quad(c)\,
tropical limit of $\log \tilde{\Omega}_0$}
\label{fig:tetra}
\end{center}
\end{figure}

Note that the leading contribution can be computed as 
the volume of the tetrahedron. 
This can be easily seen from Fig.\ref{fig:tetra}.
\begin{align}
K_w^{trop}&=c \cdot \text{volume of the tetrahedron} \notag \\
&=
\frac{c}{24\Delta_{[123]} }\frac{1}
{\alpha_1 \alpha_2 \alpha_3}(\log|\phi|^2)^3.
\end{align}
It is important to note that there are no subleading 
terms of the form $(\log |\phi|^2)^2$, as shown in 
Eq.(\ref{eq:asymptotic_behavior}) in 
Appendix~\ref{sc:gp}. 
Since deviations from the tropical limit are localized in finite 
regions around walls and junctions, 
there are contributions 
proportional to the wall length $\log |\phi|^2$  
and constant contributions associated with junctions. 
Therefore
there is no contribution 
proportional to the area $(\log |\phi|^2)^2$ in Fig.\ref{fig:tetra}. 
This tropical approximation should be valid for 
finite values of coupling constant $g$. 

\begin{figure}
\begin{center}
\includegraphics[width=70mm]{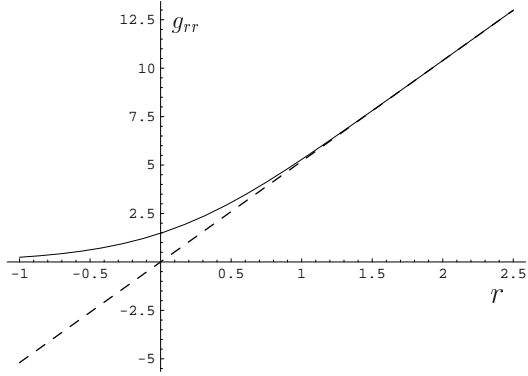}
\caption{Metric of moduli space (solid line) 
$g_{rr}$ and asymptotic metric (dashed line). 
\hs{25}We have chosen 
$\vec m_1 = (1,~0), 
~\vec m_2=(-1/2,~\sqrt{3}/2), ~\vec m_3=(-1/2,~-\sqrt{3}/2)$, 
and $c=1,\,g\to\infty$.}
\label{fig:metric3}
\end{center}
\end{figure}

The variation of this potential yields the following 
simple asymptotic metric:
\begin{align}
ds_w^2&=
\frac{c}{\Delta_{[123]}}\frac{r}{\alpha_1 \alpha_2 \alpha_3}
(dr^2+d\theta^2) \notag \\
&=
\frac{c}{2\Delta_{[123]}}\frac{1}{\alpha_1 \alpha_2 \alpha_3}
\left(dR^2+\left(\frac{3}{2} R\right)^{\frac{2}{3}} d\theta^2\right), 
\label{eq:metw}
\end{align}
where subleading contributions 
should be suppressed by powers of $1/|\phi|\sim e^{-r}$, 
as illustrated in Appendix \ref{appendix:correction} 
for the case of walls.
In Fig.\ref{fig:metric3}, the numerically evaluated metric $g_{rr}$ of 
moduli space is compared to the asymptotic metric (\ref{eq:metw}) 
evaluated at $r\to \infty$ with $g\to \infty$.

\subsection{Tropical limit at Finite Couplings}\label{subsec:finite}

In the previous subsection, we have evaluated the asymptotic behavior for large values of moduli and obtained the K\"ahler potential in the tropical limit.
This tropical limit contains only informations of vacua 
and does not require the detailed informations on the 
internal structure of walls and/or junctions. 
The tropical limit
should be applicable to a solution obtained in 
arbitrary coupling constant. 
In the following we use vacuum expectation values
to evaluate Eq.~(\ref{eq:kah_g}).

The key point is that from the second equation of Eq.(\ref{eq:solBPS}) 
$\Sigma_\alpha$ can be written as
\begin{align}
\Sigma_\alpha=\frac{1}{2}S^{-1}\partial_\alpha \Omega S^{\dag -1}.
\end{align}
Using this relation,  
the integrand in Eq.(\ref{eq:kah_g}) can be rewritten as 
\begin{eqnarray}
\frac{1}{2g^2}\Tr \bigl[ (\Omega^{-1}\partial_\alpha \Omega)^2 \bigr]
=\frac{2}{g^2}\Tr~\Sigma_\alpha^2.
\label{eq:second}
\end{eqnarray}
Note that in the vacuum $\langle A \rangle$, 
$\Sigma_\alpha^2$ takes the value 
\begin{align}
\langle \Sigma_\alpha^2 \rangle_{\langle A \rangle}=m_A^2+n_A^2.
\end{align}
See Eq.(\ref{eq:sigma}). Therefore, the tropical limit yields 
\begin{align}
\frac{1}{2g^2}\Tr \bigl[ (\Omega^{-1}\partial_\alpha \Omega)^2 \bigr]
\rightarrow \frac{2}{g^2}|\vec m_A|^2
\end{align}
in region $\langle A \rangle$. 
Note that this is constant and does not depend on the 
coordinates $x^1$ and $x^2$.

\begin{figure}[h]
\begin{center}
\begin{tabular}{ccc}
\includegraphics[width=5cm]{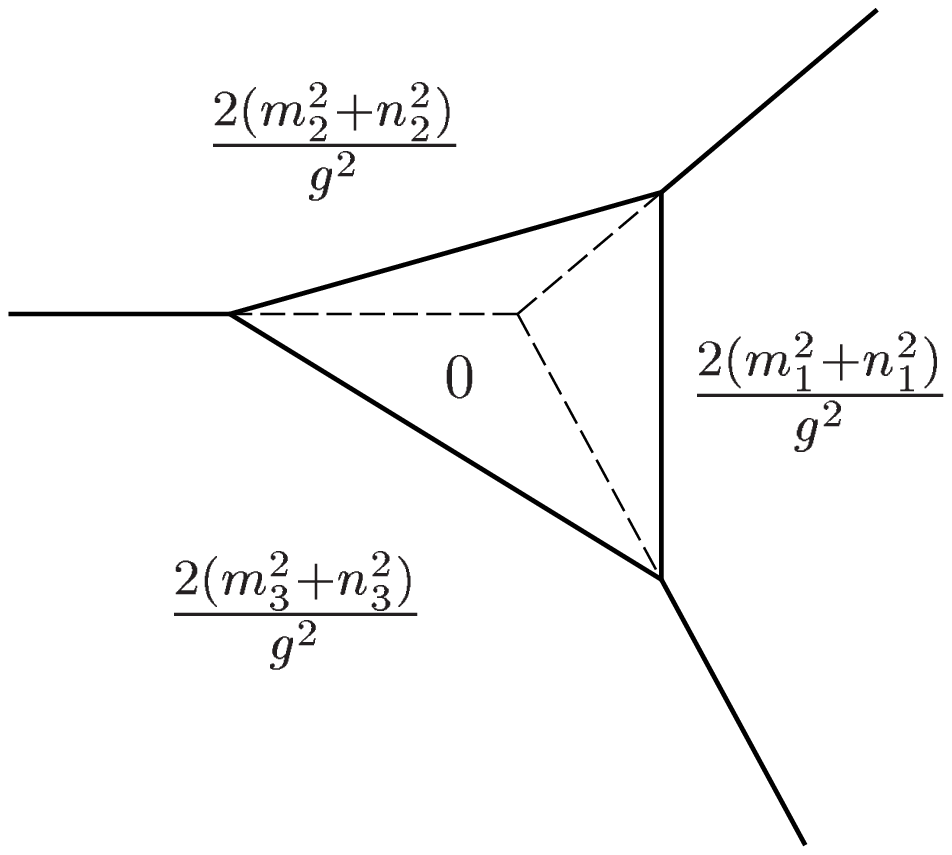}&
\includegraphics[width=5cm]{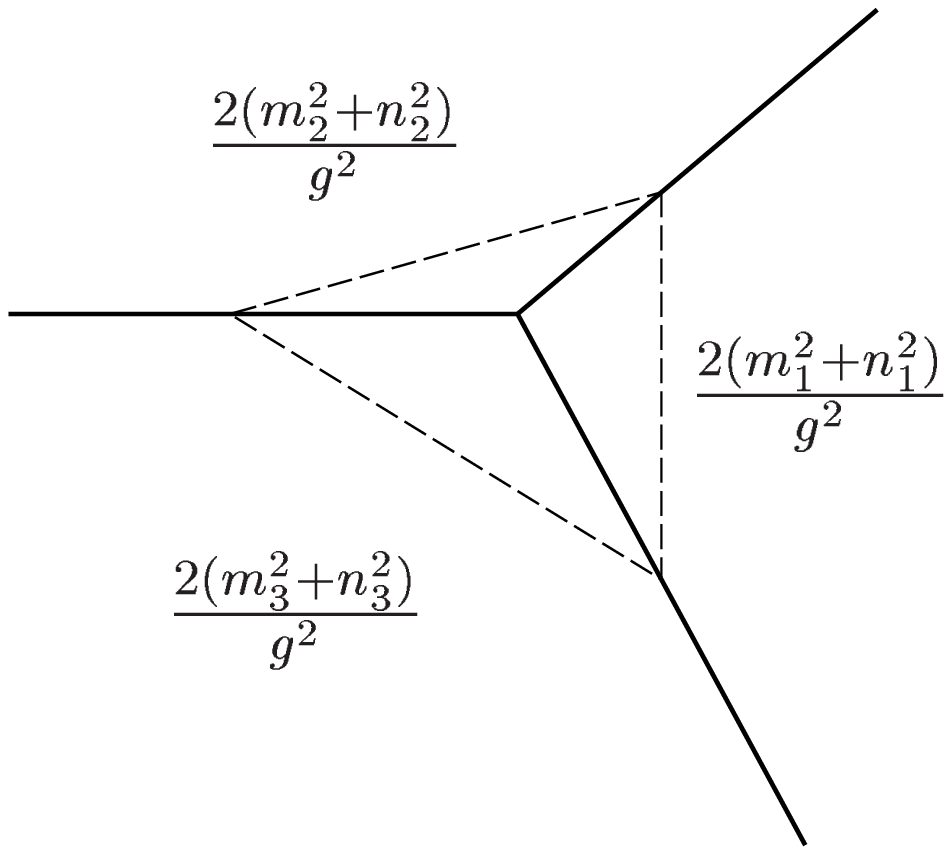}&
\includegraphics[width=5cm]{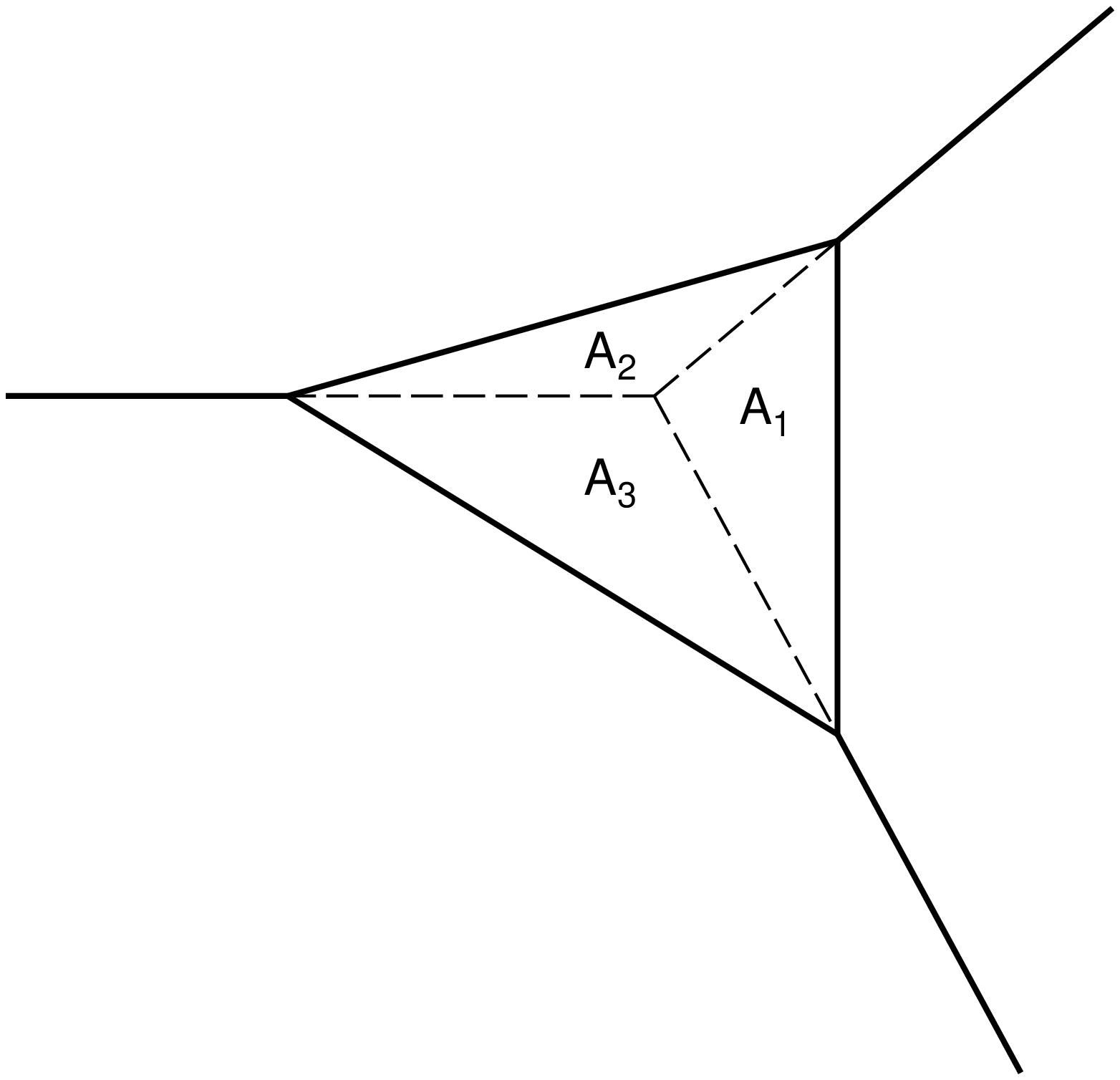}\\
(a)&(b)&(c)
\end{tabular}
\caption{(a)\,tropical limit of $\frac{1}{2g^2}\text{Tr}\bigl[(\Omega^{-1}\partial_\alpha\Omega)^2\bigr],$\quad
(b)\,tropical limit of $\frac{1}{2g^2}\text{Tr}\bigl[(\tilde{\Omega}^{-1}\partial_\alpha\tilde{\Omega})^2\bigr]$,\quad
(c)\,division of area of the loop}
\label{fig:finite}
\end{center}
\end{figure}

Before trying to compute, we remove moduli-independent 
divergence by using the K\"ahler transformation
as before 
\begin{align}
K_g \equiv \int d^2x 
\frac{1}{2g^2}\Tr \bigl[ (\Omega^{-1}\partial_\alpha \Omega)^2 
-(\tilde{\Omega}^{-1}\partial_\alpha \tilde{\Omega})^2 \bigr].
\end{align}
Here $\tilde{\Omega}$ means $\tilde{\Omega}=\Omega(|\phi|=0)$. 
In the tropical limit, the external vacuum regions 
 $\langle A \rangle, (A=1,2,3)$ give vanishing contributions, 
whereas the vacuum region $\langle 4 \rangle$ inside the loop 
gives a non-zero contribution, 
as before. 
See Fig.~\ref{fig:finite} (a) and (b).
Therefore, the tropical limit yields 
\begin{align}
K_g^{trop}&=-\frac{2}{g^2}|\vec m_1|^2A_1-\frac{2}{g^2}|\vec m_2|^2A_2-\frac{2}{g^2}|\vec m_3|^2A_3 \notag \\
&=
-\frac{1}{4g^2\Delta_{[123]}}\left(\frac{|\vec m_{12}|^2}{\alpha_3}+\frac{|\vec m_{23}|^2}{\alpha_1}+\frac{|\vec m_{31}|^2}{\alpha_2}\right)(\log |\phi|^2)^2.
\end{align}
Here $A_i$ is 
the area 
of the loop, which is given in 
Fig.\ref{fig:finite}(c) and $\vec m_{AB} = \vec m_A - \vec m_B$. 
The variation of this potential leads to the following 
contribution to the metric.
\begin{align}
ds_g^2=
-\frac{1}{g^2\Delta_{[123]}}\left(\frac{|\vec m_{12}|^2}{\alpha_3}+\frac{|\vec m_{23}|^2}{\alpha_1}+\frac{|\vec m_{31}|^2}{\alpha_2}\right)(dr^2+d\theta^2).\label{eq:metg}
\end{align}

Let us combine the contributions $ds_w^2$ 
in Eq.(\ref{eq:metw}) 
and $ds_g^2$ 
in Eq.(\ref{eq:metg}) together 
to give the total K\"ahler metric in the asymptotic 
region obtained by the tropical limit 
\begin{align}
ds^2=&
\frac{c}{\Delta_{[123]}}\left[ \frac{r}{\alpha_1 \alpha_2 \alpha_3}
-\frac{1}{g^2 c}\left(\frac{|\vec m_{12}|^2}{\alpha_3}+\frac{|\vec m_{23}|^2}{\alpha_1}+\frac{|\vec m_{31}|^2}{\alpha_2}\right)\right](dr^2+d\theta^2).
\end{align}

Since the tropical limit does not require the details of 
the solution and contains only the information of vacua,
this result 
should be applicable to solutions for the general cases with 
arbitrary masses and arbitrary coupling $g$. 
Using the explicit solution in the strong coupling 
limit $\Omega=\Omega_0$, 
we will estimate the correction terms and show in Appendix B 
that the corrections to the K\"ahler metric of the tropical 
limit should be suppressed exponentially for large $r$.

\subsection{Tropical Limit in Non-Abelian Single Triangle Loop}
\label{subsec:trononabelian}

In this section, let us consider 
the non-Abelian gauge 
theory with $\NC=3$, $\NF=4$. 
This model has also four vacua, 
$\langle A_r \rangle=\langle 123 \rangle,
\langle 124 \rangle, \langle 134 \rangle$ and 
$\langle 234 \rangle$.
In order to compare with the Abelian model, we choose
the same mass parameters as 
that of the Abelian gauge 
theory with $\NC=1$, $\NF=4$ in Sec.~\ref{subsec:1-loop}. 

Although the moduli matrix $H_0$ is
$3 \times 4$ in this case, we extract $3 \times 3$ matrix 
$H_0^{\langle A_r \rangle}$ defined by 
$(H_0^{\langle A_r \rangle})^s_{~t}=(H_0)^s_{~A_t}$. 
Then we choose moduli parameters as 
\begin{eqnarray}
\det H_0^{\langle 234 \rangle}
=\det H_0^{\langle 134 \rangle}=\det H_0^{\langle 124 \rangle}=1,
\hs{5}
\det H_0^{\langle 123 \rangle}=\phi=e^{r+i\theta}.
\label{eq:NA_moduli_triangle}
\end{eqnarray}
There is an exact correspondence between vacua of the 
$U(N_{\rm C})$ gauge theory with $N_{\rm F}$ flavors to 
those of the $U(N_{\rm F}-N_{\rm C})$ gauge theory with 
$N_{\rm F}$ flavors\cite{Arai:2003tc}. 
The moduli space point (\ref{eq:NA_moduli_triangle}) 
in the $U(3)$ gauge theory and the moduli space point 
(\ref{eq:A_triangle_moduli}) in the $U(1)$ gauge theory 
give different ${\rm det}\Omega$, which reduces to the 
same ${\rm det}\Omega_0$ near vacuum regions because of 
this correspondence. 
In the tropical limit, we evaluate $\Omega$ by 
replacing it with $\Omega_0$ dominated by 
the single dominant vacuum weight. 
We only need informations on vacuum regions. 
Therefore, the asymptotic metric (\ref{eq:metw}) 
corresponding to the 
first term in Eq.(\ref{eq:kah}) is also 
valid in the $U(3)$ gauge theory. 

The difference appears 
in 
the second term in Eq.(\ref{eq:kah}). 
From Eq.(\ref{eq:second}), the tropical limit leads to 
\begin{align}
\frac{1}{2g^2}\Tr \bigl[ (\Omega^{-1}\partial_\alpha \Omega)^2 \bigr]
\rightarrow \frac{2}{g^2}(|\vec m_A|^2+|\vec m_B|^2+|\vec m_C|^2)
\end{align}
in 
the vacuum $\langle ABC \rangle$ of the $U(3)$ gauge theory 
corresponding to the $\langle D \rangle$ of the $U(1)$ 
gauge theory. 
If we remove the moduli-independent divergence as before,
non-zero contribution comes only from the vacuum 
$\langle 123 \rangle$.
Therefore, this yields the same form of the
asymptotic metric but its sign is flipped
\begin{align}
ds_g^2=
\frac{1}{2g^2\Delta_{[123]}}\Bigl(\frac{|\vec m_{12}|^2}{\alpha_3}
+\frac{|\vec m_{23}|^2}{\alpha_1}
+\frac{|\vec m_{31}|^2}{\alpha_2}\Bigr)(dr^2+d\theta^2).
\label{eq:metg-non}
\end{align}
This 
result is quite intriguing. 
The meaning of the plus sign will be clarified in 
section \ref{sec:understand}.

\section{Understanding of the Asymptotic Metric}
\label{sec:understand}

In this section we clarify the physical meaning of the 
asymptotic metric. 

\subsection{Kinetic Energy of Walls}\label{subsec:kinetic}

We are now considering the dynamics of a system which 
consists of three internal walls and has two kinematic 
variables $r$ and $\theta$. 
(Three external walls have been fixed by 
boundary conditions.
) 
There is no potential 
energy 
among the three walls due to the BPS 
nature of this system. Therefore, the effective 
Lagrangian of single triangle loop would
be simply given by the kinetic energy of internal walls.

The metric (\ref{eq:metw}) expresses this kinetic energy 
of walls. 
Let us confirm this statement. \\
First of all, lengths of internal walls are given by
\begin{align}
\langle 1 \rangle \langle 4 \rangle ~\text{wall}\quad
l^{(1,4)} = \frac{|{\vec m_1}|}{\Delta_{[123]}}\frac{1}{\alpha_3 \alpha_2}r, \notag \\
\langle 2 \rangle \langle 4 \rangle ~\text{wall}\quad
l^{(2,4)} = \frac{|{\vec m_2}|}{\Delta_{[123]}}\frac{1}{\alpha_1 \alpha_3}r, \notag \\
\langle 3 \rangle \langle 4 \rangle ~\text{wall}\quad
l^{(3,4)} = \frac{|{\vec m_3}|}{\Delta_{[123]}}\frac{1}{\alpha_2 \alpha_1}r.
\end{align}
Tensions of internal walls are
\begin{align}
\langle 1 \rangle \langle 4 \rangle ~\text{wall}\quad T^{(1,4)}=c|\vec m_1|,\notag \\
\langle 2 \rangle \langle 4 \rangle ~\text{wall}\quad T^{(2,4)}=c|\vec m_2|\notag, \\
\langle 3 \rangle \langle 4 \rangle ~\text{wall}\quad T^{(3,4)}=c|\vec m_3|.
\end{align}
Next we have to know how the internal walls move when the variable $r$ changes. We denote the distances between the origin
and internal walls as $l_0^{(A4)}, (A=1,2,3)$. These are given by 
\begin{align}
\langle 1 \rangle \langle 4 \rangle ~\text{wall}\quad l_0^{(1,4)}=\frac{r}{|\vec m_1|},\notag \\
\langle 2 \rangle \langle 4 \rangle ~\text{wall}\quad l_0^{(2,4)}=\frac{r}{|\vec m_2|},\notag \\
\langle 3 \rangle \langle 4 \rangle ~\text{wall}\quad l_0^{(3,4)}=\frac{r}{|\vec m_3|}.
\end{align}
Therefore, the 
total 
kinetic energy of internal walls is 
\begin{align}
T=&\sum_{A=1}^3 \frac{1}{2}T^{(A,4)} l^{(A,4)}\left(\frac{d}{dt}l^{(A,4)}_0\right)^2 
=\frac{c}{2\Delta_{[123]}}\frac{r}
{\alpha_1\alpha_2\alpha_3}\left(\frac{dr}{dt}\right)^2.
\end{align}
We observe that the moduli space metric at $g^2\to \infty$ 
in Eq.(\ref{eq:metg}) 
agrees with this result, since the total kinetic energy 
should be expressed in terms of the metric as 
\begin{equation}
T=\frac{1}{2} \left(\frac{ds_w}{dt}\right)^2. 
\end{equation}
\subsection{Kinetic Energy of Junction Charges}
\label{subsec:junction}
In the above, we considered only the kinetic energy of 
internal walls. 
However, the mass of a wall web actually consists of 
energies 
of walls (tension $T_w$ times the length of the walls) 
and a contribution from the junction charge $Y$. 
The $Y$ contributes negatively to the energy in the 
$U(1)$ gauge theory and is interpreted as the 
binding energy of the three walls. 
If $r$ changes to $r+dr$, the junction points also moves 
slightly, which accompanies the movement of junction 
charges at the junction points.
Therefore, it is natural to expect that the metric 
(\ref{eq:metg}) 
expresses the ``kinetic energy'' 
of junction charges, which is  negative because of the 
negative ``mass'' of the junction. 
Junction charges are given by
\begin{align}
\langle 124 \rangle ~\text{junction}\quad Y^{124}=
-\frac{\Delta_{[123]}}{g^2}\alpha_3,\notag \\
\langle 234 \rangle ~\text{junction}\quad Y^{234}=
-\frac{\Delta_{[123]}}{g^2}\alpha_1, \notag \\
\langle 314 \rangle ~\text{junction}\quad Y^{314}=
-\frac{\Delta_{[123]}}{g^2}\alpha_2.
\label{eq:junction}
\end{align}
Distances between the origin and the junctions are
\begin{align}
\langle 124 \rangle ~\text{junction}\quad 
l^{124}=\frac{|\vec m_{12}|}{\Delta_{[123]}\alpha_3}r, \notag \\
\langle 234 \rangle ~\text{junction}\quad 
l^{234}=\frac{|\vec m_{23}|}{\Delta_{[123]}\alpha_1}r, \notag \\
\langle 314 \rangle ~\text{junction}\quad 
l^{314}=\frac{|\vec m_{31}|}{\Delta_{[123]}\alpha_2}r.
\end{align}
Using these, we obtain the kinetic energy of junction charges
\beq
T&=& T^{124}+T^{234}+T^{314}\notag \\
&=&
-\frac{1}{2g^2\Delta_{[123]}}\left(\frac{|\vec m_{12}|^2}{\alpha_3}+\frac{|\vec m_{23}|^2}{\alpha_1} +\frac{|\vec m_{31}|^2}{\alpha_2} \right)
\left(\frac{dr}{dt}\right)^2.
\eeq
This agrees with the result in Eq.(\ref{eq:metg}) of the 
Abelian gauge theory. 

This interpretation naturally explains the sign 
difference between Eq.(\ref{eq:metg}) and 
Eq.(\ref{eq:metg-non}). 
In the non-Abelian junction, junction charge gives a 
positive contribution
to energy 
of a wall web in contrast to 
the Abelian junction. 
This can be interpreted as the $Y$-charge of the Hitchin 
system. 
For the details about these issues, see \cite{Eto:2005fm}.

The non-Abelian junction charges have the same magnitude as 
the Abelian junction charges (\ref{eq:junction}) 
but its sign is reversed. 
\begin{align}
\langle 234 \rangle \langle 134 \rangle \langle 123 \rangle \quad
\text{junction}\quad Y^{124}=
\frac{\Delta_{[123]}}{g^2}\alpha_3, \notag \\
\langle 134 \rangle \langle 124 \rangle \langle 123 \rangle \quad
\text{junction}\quad Y^{234}=
\frac{\Delta_{[123]}}{g^2}\alpha_1, \notag \\
\langle 124 \rangle \langle 234 \rangle \langle 123 \rangle \quad
\text{junction}\quad Y^{124}=
\frac{\Delta_{[123]}}{g^2}\alpha_2.
\end{align}
This explains the sign flip in the asymptotic metric 
of Eq.(\ref{eq:metg-non}) compared to that of 
Eq.(\ref{eq:metg}).

\section{Conclusion and Discussion}\label{sec:discussion}

We have constructed an effective theory of webs/networks of domain walls. The general form of the effective action Eq.~(\ref{eq:general}) is written in terms of the K\"ahler potential given by Eq.~(\ref{eq:kah}), (\ref{eq:kah_w}), (\ref{eq:kah_g}). In the strong coupling limit $g^2 \rightarrow \infty$, the K\"ahler potential for the single triangle loop can be written in the form of the power series as Eq.~(\ref{eq:kahlerw}). The moduli space is smooth even at the point where the size of the loop shrinks to zero. The large size behavior of the K\"ahler potential can be calculated as Eq.~(\ref{eq:metw}) by using the tropical limit. In this limit we can also evaluate the contributions from the junction charges in the Abelian case Eq.~(\ref{eq:metg}) and in the non-Abelian case Eq.~(\ref{eq:metg-non}). These asymptotic metrics can be read from kinetic energy of the constituent walls of the loop and the junction charges.

Let us summarize possible future works.
Since the metric of the moduli space of a single loop 
has been obtained, 
its dynamics can be discussed by the moduli space 
(Manton's) approximation. 
We also should extend our work to 
multiple loops of domain walls. 
These aspects will be reported elsewhere \cite{EFNNOS}.

The parallel domain walls have dyonic extension \cite{Lee:2005sv} 
if we introduce complex masses. 
In the case of domain wall webs, they admit 
dyonic extension if the triplet of masses 
is introduced \cite{Eto:2005sw}. 
In the case of supersymmetric field theory, 
this is possible in a $d=2+1$ theory
which can be obtained by the Scherk-Schwarz dimensional 
reduction from the $d=3+1$ theory discussed in this paper. 
The effective action on webs becomes a classical mechanics 
(a $d=1$ nonlinear sigma model). 
Since it is known that a potential, 
which is proportional to a square of a Killing vector 
of a $U(1)$ isometry of the moduli space, 
is induced on the effective theory of dyons \cite{Tong:1999mg}
or dyonic instantons \cite{Lambert:1999ua}, 
the same kind of a potential,  
with a $U(1)$ Killing vector associated to a loop, 
is expected on a classical mechanics 
on the moduli space of the dyonic webs of domain walls. 

As noted in the introduction, 
the present work may be applied to 
construction of 1/8 BPS solitons \cite{Lee:2005sv,Eto:2005sw} . 
Unfortunately, 
as discussed above, 
only one time dimension is left in the effective theory 
on the wall web 
when one needs a potential on the moduli space 
in the framework of supersymmetric field theory.  
Therefore one can construct 
a 1/2 BPS domain wall on the wall web
only in an Euclidean theory, 
which gives a space-like brane. 
To do this, however, a single triangle loop is not enough 
because the effective theory on it has only one vacuum 
due to the fact that 
the potential is the square of 
a $U(1)$ Killing vector associated to a loop.
One would need at least two loops
to obtain two disconnected vacua
in the effective theory on the wall web. 
Sigma model lumps do not require a potential, 
so lumps on the wall webs could be constructed 
in four space time dimensions. 
To do this, the moduli space must contain 
a non-trivial second homotopy group, 
$\pi_2 (M) \neq 0$. 
One would need at least multiple loops. 
This case of the lump solution also requires to consider 
a Euclidean theory.

\subsubsection*{Acknowledgements}

We would like to thank Kazutoshi Ohta for useful discussions 
on tropical limit and David Tong for valuable comments on 
$(2,0)$ models. 
This work is supported in part by Grant-in-Aid for 
Scientific Research from the Ministry of Education, 
Culture, Sports, Science and Technology, Japan No.17540237
and No.18204024 (N.~S.). 
The works of 
K.~O. and M.~E. 
are 
supported by Japan Society for the Promotion 
of Science under the Post-doctoral 
Research 
Program. 
T.~F. gratefully acknowledges 
support from a 21st Century COE Program at 
Tokyo Tech ``Nanometer-Scale Quantum Physics" by the 
Ministry of Education, Culture, Sports, Science 
and Technology, 
and 
support from the Iwanami Fujukai Foundation.


\appendix
\section{Hypergeometric functions}
In this appendix, we show how to express the K\"ahler 
potential $K_w$ in Eq.(\ref{eq:kahlerw}) 
in the strong coupling limit in terms of hypergeometric 
functions provided the parameters $\alpha_k$ are rational. 

\subsection{${}_pF_q\left(\{a_n\};\{b_m\};x\right)$}
\label{sc:hypergeom_func}

The hypergeometric functions are defined by the following 
power series 
\begin{eqnarray}
{}_pF_q\left(a_1,a_2,\cdots,a_p;b_1,b_2,\cdots,b_q;x\right) 
&\equiv&\sum_{k=0}^{\infty}\frac{1}{k!}\frac{(a_1)_k(a_2)_k\cdots(a_p)_k}
{(b_1)_k(b_2)_k\cdots(b_q)_k}x^k
\label{eq:hypergeom_func}
\end{eqnarray}
with  $(\nu)_k\equiv
(k+\nu-1)(k-2+\nu)\cdots(\nu)=\Gamma(k+\nu)/\Gamma(\nu)$, ($k!=(1)_k$). 

\subsection{$G_p(\{\alpha_n\};x)$}\label{sc:gp}
Let us define a function $G_p(\{\alpha_k\};x)$  
for $p=1,2,3,\cdots, $ with positive real parameters $\alpha_k,(k=1,\cdots,p)$ 
by an infinite series
\begin{eqnarray}
G_p(\alpha_1,\alpha_2,\cdots,\alpha_p;x)&\equiv& 
\sum_{n=1}^{\infty}
\frac{\Gamma(\alpha_1 n)\Gamma(\alpha_2 n)\cdots\Gamma(\alpha_pn)}
{\Gamma(n+1)}x^n. 
\label{eq:Gpfunction}
\end{eqnarray}
Here, the radius of convergence $R$ is given by
\begin{eqnarray}
 R=\lim_{n\rightarrow\infty}(n+1)\frac{\prod_{k=1}^p\Gamma(\alpha_kn)}
{\prod_{k=1}^p\Gamma(\alpha_k(n+1))}
=e^S\lim_{n\rightarrow\infty}\frac{(n+1)}{n^u}
=\left\{\begin{array}{cc}
    \infty& u<1\\
   e^S& u=1\\
  0&  u>1 
	  \end{array}\right.\quad
\end{eqnarray}
with parameters $S,u$ defined by
\begin{eqnarray}  
S\equiv-\sum_{i=1}^p\alpha_i\log\alpha_i, \quad u\equiv\sum_{i=1}^p\alpha_i.
\end{eqnarray}
where we used the formula
\begin{eqnarray}
\lim_{z\rightarrow \infty}
e^{-\alpha \log z}\frac{\Gamma(z+\alpha)}{\Gamma(z)}=1 
\end{eqnarray}
derived from 
the Stirling series. 
In this paper we are interested in the case of $u=1$ and 
set $u=1$ from now on. 
Then $G_p$ is convergent 
for $|x|\le e^S$.  
In the case where $\{\alpha_n\}$ are rational numbers 
\begin{eqnarray}
 \alpha_n=\frac{M_n}{N},(M_n,N\in {\bf N}),\quad 
\sum_{i=1}^pM_i=N,
\end{eqnarray}
the function $G_p(\{\alpha_n\};x)$ can be rewritten by a 
finite sum of 
 hypergeometric functions $F$ defined in 
Eq.(\ref{eq:hypergeom_func}) as 
\begin{eqnarray}
 G_p(\alpha_1,\cdots,\alpha_p;x)&=&\sum_{m=1}^{N}\sum_{k=0}^\infty (n\rightarrow k N+m)\notag \\
&=&\sum_{m=1}^{N}
\frac{\Gamma(\alpha_1 m)\Gamma(\alpha_2 m)\cdots\Gamma(\alpha_p m)}{\Gamma(m+1)}x^{m} \times \notag \\
&&\quad\times
{}_{N+1}F_N
\left(1,\{A_{M_1}\},\cdots, \{A_{M_p}\};\{B_{N}\};
e^{-N S}x^N\right),\notag \\
\end{eqnarray} 
with a set of parameters given by 
\begin{eqnarray}
&A_{M_i}=\frac{m}N,\frac{m}N+\frac{1}{M_i}, \cdots, \frac{m}N+1-\frac{1}{M_i},&\notag \\
&B_N=\frac{m+1}N,\frac{m+2}N,\cdots,\frac{m+N}N,&\notag \\
&1+\sum_{i=1}^p\sum_{l=1}^{M_i}(A_{M_i})_l-\sum_{l=1}^N(B_N)_l=\frac{1-p}2,& 
\end{eqnarray}
where we used the following identity, with  $k,l\in {\bf N}$,
\begin{eqnarray}
 \Gamma(k l+\nu)&=&(k l +\nu-1)\cdots(k l+\nu-l)\notag \\
&&\times ((k-1) l +\nu-1)\cdots((k-1) l+\nu-l)\times \cdots\notag \\
&&\times ( l+\nu-1)\cdots \nu \times \Gamma(\nu)\notag \\
&=&l^{kl}\left(k  +\frac{\nu-1}l\right)\cdots
\left(k  +\frac{\nu-l}l\right)\notag \\
&&\times \left(k-1 +\frac{\nu-1}l\right)\cdots
\left(k-1  +\frac{\nu-l}l\right)\times \cdots\notag \\
&&\times \left(1  +\frac{\nu-1}l\right)\cdots \frac{\nu}l \times \Gamma(\nu)\notag \\
&=&(l^l)^k\Gamma(\nu)\prod_{i=0}^{l-1}\left(\frac{\nu+i}{l}\right)_k.
\end{eqnarray}


By making an analytic continuation of an integral 
representation of the power series expression 
(\ref{eq:Gpfunction}) 
with the similar method as in 
Sec.\ref{sec:tropical}, 
one obtains for $x\gg 1$ 
\begin{eqnarray}
-G_1(1;-x)&=&\log(1+x)\approx \log x+\frac1x-\frac1{2x^2}+\cdots,\notag \\
 -G_2(\alpha,\beta;-x)&\approx& \frac{1}{2\alpha\beta}(\log x)^2
+\left(\frac{1}{\alpha\beta}-1\right)\frac{\pi^2}{6}+{\cal O}(x^{-1}),
\notag \\
-G_3(\alpha_1,\alpha_2,\alpha_3;-x)&\approx&
\frac1{6\alpha_1\alpha_2\alpha_3}(\log x)^3+A(\{\alpha\})\log x
+B(\{\alpha\})+\dots, 
\label{eq:asymptotic_behavior}
\end{eqnarray}
where the coefficients $A$ and $B$ are given by 
\begin{eqnarray}
 A(\alpha_1,\alpha_2,\alpha_3)&=&
\frac{\pi^2}{6}\left(\frac{1}{\alpha_1\alpha_2\alpha_3}
-\frac1\alpha_1-\frac1\alpha_2-\frac1\alpha_3\right), \nn
B(\alpha_1,\alpha_2,\alpha_3)&=&\zeta(3)
\left(\frac1\alpha_1+\frac1\alpha_2+\frac1\alpha_3-1\right).
\end{eqnarray}
The coefficient $\zeta(3)$ is verified numerically to order 
$10^{-100}$. 

More generally, the leading contribution is given by 
\begin{eqnarray}
 -G_p(\{\alpha_n\};-x)&\approx&\frac{1}{p! \prod_{i=1}^p\alpha_i}(\log x)^p
+{\cal O}\left((\log x)^{p-2}\right). 
\end{eqnarray}

\section{Correction Terms}\label{appendix:correction}
The exponentially suppressed term in the K\"ahler 
potential in Eq.~(\ref{eq:Kahler}) can be calculated 
as follows:
\begin{eqnarray}
 \int_{I_1}dy \log\left(1+\frac{e_2}{e_1+e_0}\right)&=&
\sum_{n=1}^\infty\frac{(-1)^{n-1}}{n}
\int _{I_1}dy\left(\frac{e_2}{e_1+e_0}\right)^n\nn
&=&\sum_{n=1,k=0}^{\infty}(-1)^{n+k-1}\frac{
{(n+k-1)!}}{n!k!}e_0^k \int_{I_1}dy\, e_2^n\, e_1^{-(n+k)}\nn
&=&\frac1{2m}\sum_{n=1,k=0}^{\infty}(-1)^{n+k-1}\frac{
{(n+k-1)!}}{n!k!(n+k\alpha_1)}e^{-n \frac{2}{\alpha_1}r}\nn
\int_{I_2}dy \log\left(1+\frac{e_2}{e_1+e_0}\right)&=&
\sum_{n=1,k=0}^{\infty}(-1)^{n+k-1}\frac{
{(n+k-1)!}}{n!k!}
e_0^{-(n+k)} \int_{I_2}dy\, e_2^n\, e_1^{k}\nn
&=&\frac{1}{2m}\sum_{n=1,k=0}^{\infty}(-1)^{n+k-1}\frac{
{(n+k-1)!}}{n!k!(k\alpha_1-n\alpha_2)}e^{-n \frac{2}{\alpha_1}r}\nn
&&{}-\frac{1}{2m}\sum_{n=1,k=0}^{\infty}(-1)^{n+k-1}\frac{
{(n+k-1)!}}{n!k!(k\alpha_1-n\alpha_2)}e^{-2(n+k)r}\nn
\end{eqnarray}
where we assumed, to facilitate the computation, that 
$\alpha_1$ is irrational. 
The sum 
of terms proportional to integer powers of $e^{-2r}$, like 
$e^{-2 n r}$ and $e^{-2 (n+k)r}$, turns out to vanish.  
The rest give 
\begin{eqnarray}
&& \frac{1}{2m}\sum_{n=1,k=0}^{\infty}(-1)^{n+k-1}\frac{
{(n+k-1)!}}{n!k!}\left(\frac1{k\alpha_1+n}+\frac1{k\alpha_1-n\alpha_2}\right)
e^{-n \frac{2}{\alpha_1}r}+(\alpha_1\leftrightarrow \alpha_2)\nn
&=&\frac{1}{2m}\frac{\pi}{\alpha_1\sin\left(\frac{\pi}{\alpha_1}\right)}
e^{-\frac{2}{\alpha_1}r}
+{\cal O}\left(e^{-\frac{4}{\alpha_1}r}\right)
+(\alpha_1\leftrightarrow \alpha_2) . 
\end{eqnarray}
Although the above result is given under the assumption that 
 $\alpha_1(=1-\alpha_2)$ is irrational, we can confirm directly 
that the leading term of the above is also finite with rational $\alpha_1$,
especially a limit of $\alpha_1\rightarrow \frac12$
gives 
\begin{eqnarray}
\frac1{2m}\frac{\pi}{\alpha_1\sin\left(\frac{\pi}{\alpha_1}\right)}
e^{-\frac{2}{\alpha_1}r}
+(\alpha_1\leftrightarrow \alpha_2)
 &\rightarrow& -4 r e^{-4r}.
\end{eqnarray}
This result is in accord with the known 
result\cite{Isozumi:2003rp,Tong:2002hi}.


\end{document}